\documentclass[11pt,a4paper,notoc,twoside]{JHEP3}
\usepackage{graphicx,epsfig,rotating}
\usepackage{amssymb}
\usepackage{latexsym}
\usepackage{citesort}
\DeclareSymbolFontAlphabet{\mathbb}{AMSb}

\title{ The Classification of the Simply Laced Berger Graphs from
 Calabi-Yau $CY_3$ spaces}
\preprint{\small  CERN-PH-TH/2004-123\\
IFT-UAM/CSIC-04-112\\
UM-FT/04-111}
\author{ J. Ellis{}$^{1\star}$, E. Torrente-Lujan${}^{2\star}$,  G. G. Volkov${}^{1,3\star}$\\
{\small
$^1${\it TH Division, Physics Department, CERN, CH-1211 Geneva 23, 
Switzerland}\\
$^2${\it GFT, Dept. of Physics, Universidad de Murcia,Spain}\\
$^3${\it IFT, Univ. Autonoma de Madrid, Cantoblanco, Madrid, Spain, \\ on leave from PNPI, Gatchina, St 
Petersburg,Russia}
}\\
\email{ john.ellis@cern.ch},\email{ e.torrente@cern.ch},\email{ guennadi.volkov@cern.ch}}
\abstract{
The algebraic approach to the construction of the reflexive polyhedra that
yield Calabi-Yau spaces in three or more complex dimensions with K3 fibres
reveals graphs that include and generalize the Dynkin diagrams associated
with gauge symmetries.  In this work we continue to study the structure of
graphs obtained from $CY_3$ reflexive polyhedra. 
The objective is to describe the ``simply laced'' cases, those graphs 
obtained from three dimensional spaces with K3 fibers which lead 
to symmetric matrices. We study both the affine and, derived from them, 
non-affine cases.
We present root and weight structurea for them.  
We study in  particular those graphs leading to generalizations
of the exceptional simply laced cases $E_{6,7,8}$ and $E_{6,7,8}^{(1)}$. 
We show how these 
 integral matrices can be assigned:  
they  may be obtained by
relaxing the restrictions on the individual entries of the generalized
Cartan matrices associated with the Dynkin diagrams that characterize
Cartan-Lie and affine Kac-Moody algebras.  These graphs keep, however, the
affine structure present in Kac-Moody Dynkin diagrams.  
We conjecture that these
generalized simply laced graphs and associated link matrices 
may characterize
generalizations of Cartan-Lie and affine Kac-Moody algebras. 
}

\keywords{}

\begin{document}

\section{Introduction}

Progress in fundamental physics is dependent on the identification of
underlying symmetries such as general coordinate invariance or gauge
invariance. The final objective of this work is to look for possible
symmetries beyond those of the Standard Model. The latter is based on
Cartan-Lie Algebras and their direct products, and is very successful.
there have been valiant efforts to extend the Standard Model within the
framework of Cartan-Lie algebras and with the objective of, for example,
reducing the number of free parameters appearing in the theory. however,
attempts to formulate Grand Unified theories in which the direct product
of the symmetries of the Standard Model is embedded in some larger simple
Cartan-Lie group have not had the same degree of success as the Standard
Model. The alternative possibility of unifying the gauge interactions with
gravity in some `Theory of Everything' based on string theory is very
enticing, in particular because this offers novel algebraic structures.

At a very basic level, and without any obvious direct interest for the
content of the Standard Model, Cartan-Lie symmetries are closely connected
to the geometry of symmetric homogeneous spaces, which were classified by
Cartan himself. Subsequently, an alternative geometry of non-symmetric
spaces appeared, and their classification was suggested in 1955 by Berger
using holonomy theory~\cite{Berger}. There are several infinite series of
spaces with holonomy groups $SO(n)$, $U(n)$, $SU(n)$, $Sp(n)$ and
$Sp(n)\times Sp(1)$, and additionally some exceptional spaces with
holonomy groups $G(2)$, $Spin(7)$, $Spin(16)$.

Superstring theories offer new clues how to attack the problem of the
nature of symmetries at a very basic geometric level.  For example, the
compactification of the heterotic string leads to the classification of
states in a representation of the Kac-Moody algebra of the gauge group
$E_8\times E_8$ or $Spin(32)/{ Z}_2$. These structures arose in
compactifications of the heterotic superstring on 6-dimensional Calabi-Yau
spaces, non-symmetric spaces with an $SU(3)$ holonomy group~\cite{CHSW}.
It has been shown~\cite{belavin} that group theory and algebraic
structures play basic roles in the generic two-dimensional conformal field
theories (CFTs) that underlie string theory. The basic ingredients here
are the central extensions of infinite-dimensional Kac-Moody algebras.
There is a clear connection between these algebraic and geometric
generalizations. Affine Kac-Moody algebras are realized as the central
extensions of loop algebras, namely the sets of mappings on a compact
manifold such as $S^1$ that take values on a finite-dimensional Lie
algebra. Superstring theory contains a number of other
infinite-dimensional algebraic symmetries such as the Virasoro algebra
associated with conformal invariance and generalizations of Kac-Moody
algebras themselves, such as hyperbolic and Borcherd algebras.

In connection with Calabi-Yau spaces, (Coxeter-)Dynkin diagrams which are
in one-to-one correspondence with both Cartan-Lie and Kac-Moody algebras
have been revealed through the technique of 
 the crepant
resolution of specific quotient singular structures such as the
Kleinian-Du-Val singularities ${ {\bf C}^2/G}$~\cite{DuVal}, where $G$ is
a discrete subgroup of $SU(2)$.
 Thus,
the rich singularity structure of some examples of non-symmetrical
Calabi-Yau spaces provides another opportunity to uncover
infinite-dimensional affine Kac-Moody symmetries.  The Cartan matrices of
affine Kac-Moody groups are identified with the intersection matrices of
the unions of the complex proyective lines resulting from the blow-ups of
the singularities. For example, the crepant resolution of the
${ {\bf C}^2/Z_n}$ singularity gives for rational, i.e., genus-zero, (-2)
curves an intersection matrix that coincides with the A$_{n-1}$ Cartan
matrix.
This is also  the case of $K3 \equiv CY_2$
spaces, where the classification of the degenerations of their 
elliptic 
fibers 
(which can be written in Weierstrass form)
and
their associated singularities leads to a link between $CY_2$ spaces and
the infinite and exceptional series of affine Kac-Moody algebras,
$A_r^{(1)}$, $D_{r}^{(1)}$, $E_6^{(1)}$, $E_7^{(1)}$ and $E_8^{(1)}$ (ADE)~\cite{Kodaira,Ber}.

The study of the Calabi-Yau spaces appearing in superstring, F and M
theories can be approached via the theory of toric geometry and the
Batyrev construction~\cite{Bat} using reflexive polyhedra. The concept of
reflexivity or mirror symmetry has been linked ~\cite{CF} to the problem
of the duality between superstring theories compactified on different $K3$
and $CY_3$ spaces. The same Batyrev construction has also been used to
show how subsets of points in these reflexive polyhedra can be identified
with the Dynkin diagrams~\cite{CF,CPR,Greene,KV} of the affine versions of
the gauge groups appearing in superstring and F-Theory. More explicitly,
the gauge content of the compactified theory can be read off from the {\em
dual} reflexive polyhedron of the Calabi-Yau space which is used for the
compactification.

In the case of a $K3 = CY_2$ Calabi-Yau space, any subdivision of the
reflexive polyhedron into different subsets separated by a polygon which
is itself reflexive is equivalent to establishing a fibration structure
for the space, whose fiber is simply being the space corresponding to the
intermediate mirror polygon. For example, a reflexive polyhedron
intersected by a plane yields a planar reflexive polygon separating the
`top' and `bottom' subsets in the nomenclature of~\cite{CF}, called `left'
and `right' in this work. Subsets of points in these reflexive 
polyhedra are those which can be identified
with the Dynkin diagrams~\cite{CF,CPR,Greene,KV} of the affine Kac-Moody 
algebras. It is however necessary to stress that this task was 
facilitated by the  a priori knowledge of the fiber structure, the 
reflexive Weierstrass triangle in those cases. 
Until the emergence of the UCYA, 
the absence of a systematic way of determining the slice structure 
in  generic $CY_n$ has prevented  further progress in this 
area and the finding of new Dynkin or generalized Dynkin diagrams.

Since Calabi-Yau spaces may be characterized geometrically by reflexive
Newton polyhedra, they can be enumerated systematically~\cite{Skarke}.
Moreover, one can beyond simple enumeration, as it has been recently
realized that different reflexive polyhedra are related algebraically via
what has been termed the Universal Calabi-Yau Algebra (UCYA). The term
`Universal' is motivated by the fact that it includes ternary and
higher-order operations, as well as familiar, beyond binary operations.

The UCYA is particularly well suited for exploring the fibrations of
Calabi-Yau spaces, which are visible as lower-dimensional slice or
projection structures in the original polyhedra. 
The knowledge of the  slice structure (see table (1) in this work 
of some illustration) allows us to 
uncover and understand not only  Dynkin structures in K3 and elliptic polyhedra
but new graphs in $CY_n$ polyhedra.
For an example of an
elliptic fibration of a K3 space, see Fig.~(1) in Ref.\cite{Vol} and its
accompanying description. The `left' and `right' parts of this reflexive
polyhedron both correspond to so-called `extended vectors'.
 In the
UCYA scheme, the binary operation of summing these two extended vectors
gives a true reflexive vector, that characterizes the full $CY_2 =K3$
manifold.  establishing a direct algebraic relation between K3(= CY$_2$)
and CY$_3$ spaces. This property is completely general: it has been shown
previously how the UCYA, with its rich structure of binary and
higher-order operations, can be used to generate and interrelate $CY_n$
spaces of any order. The UCYA provides a complete and systematic
description of the analogous decompositions or nestings of fibrations in
Calabi-Yau spaces of any dimension.

One of the remarkable features of Fig.~(1) in Ref.\cite{Vol} is that the right
and left sets of nodes constitute graphs corresponding to affine Dynkin
diagrams: namely the $E_6^{(1)}$ and $E^{(1)}_8$ diagrams.  This is not a
mere coincidence or an isolated example. As discussed there,
in Ref.\cite{Vol}, all the elliptic fibrations of K3 spaces found using the
UCYA construction feature this decomposition into a pair of graphs that
can be interpreted as Dynkin diagrams.

The purpose of this paper is to continue the work
already initiated in Refs.\cite{Vol,volemi}
 on the generalization of the previous results for K3
spaces to Calabi-Yau spaces in any dimension and with any fiber structure.
The main objective here is to describe the ``simply laced'' cases, those graphs 
obtained from three dimensional spaces with K3 fibers which lead 
to symmetric matrices.
As was first shown
in Ref.\cite{Vol}, many new diagrams - which we term `Berger Graphs' - can be
found in this way. 
In Ref.\cite{volemi} we gave a more formal and comprehensive 
 definition of Berger graphs and matrices. Some examples of 
planar and non-planar 
diagrams obtained from CY3 were presented and studied. It was seen there 
how some of those diagrams could be extended into infinite series 
while some others could be considered exceptional, not extendable.
We hypothesize that Berger graphs correspond, in some
manner that remains to be defined, to some new algebraic structure, just
as Dynkin diagrams are in one-to-one correspondence with root systems and
Cartan matrices in semi-simple Lie Algebras and affine Kac-Moody algebras.

Our final  objective would be  to construct a theory similar to Kac-Moody
algebras, in which newly extensions of Cartan matrices fulfilling
generalized conditions are introduced.  There are plenty of possible
generalizations of Cartan matrices obtainable by modifying the rules for
the diagonal and off-diagonal entries in the matrices, and it is
impossible to find all of them and classify them. On the other hand,
probably not all of them give meaningful, consistent generalizations of
Kac-Moody algebras, and probably fewer of them have interesting
implications for physics.  One has to find natural conditions on these
matrices, hopefully inspired by physics. The relation of Berger Graphs to
Calabi-Yau spaces could be this inspirational physical link. Once one has
the equivalent of the Cartan matrix, one can use standard algebraic tools,
such as the definition of an inner product, the construction of a root
system, its group of transformations, etc., which could be helpful in
clarifying the meaning and significance of this construction.

The structure of this paper is as follows. In section 2 we show how to
extract graphs directly from the polyhedra associated with Calabi-Yau
spaces and how one can define new, related graphs by adding or removing
nodes. 
We make the important remark that this is possible in the 
UCYA formalism because because of its ability to give naturally the 
complex slice structure of the Calabi-Yau spaces.
In section 3 we present a review of the formal algebraic definition 
of Berger graphs and matrices.
In section 4 we present a list of simply laced graphs obtained 
from CY3 spaces and give a general description of their properties.
On continuation we illustrate these properties 
 in some more detail with some examples.
Finally, in section 4 we draw some conclusions and make some conjectures.

\section{UCYA and generalized Dynkin diagrams.}

One of the main results in the Universal Calabi-Yau Algebra (UCYA)  is
that the reflexive weight vectors (RWVs) $\vec{k_n}$ of dimension $n$,
which are the fundament for the construction of CY spaces,
 can
obtained directly from lower-dimensional RWVs 
$\vec{k_{1}}, \ldots, \vec{k}_{n-r+1}$ by algebraic constructions of arity
$r$~\cite{AENV1,AENV2,AENV3,AENV4}. The dimension of the corresponding 
vector is $d+2$ for a Calabi-Yau $CY_d$ space.

For example, the sum of vectors,  
a binary composition rule of the UCYA, gives complete information about
the $(d-1)$-dimensional slice structure of $CY_d$ spaces.
In the K3 case, the 
Weierstrass fibered 91 reflexive weight vectors of the total of 95  $\vec{k_4}$ can be 
obtained by such binary, or arity-2, constructions out of just 
five RWVs of dimensions 1,2 and 3.

In an iterative process, we can combine by the same 2-ary operation 
the five vectors of dimension $1,2,3$ 
with these other 95 vectors 
to obtain a set of 4242 chains of 
 five-dimensional RWVs $\vec{k_5}$ 
CY$_3$ chains.
This process is  summarized
in Fig.~(3) in Ref.\cite{Vol}.
By construction, the corresponding mirror 
CY$_3$ spaces are shown to possess K3 fiber
bundles.
 In this case, reflexive 4-dimensional polyhedra
are also separated into three parts: a reflexive 3-dimensional
intersection polyhedron and `left' and `right' skeleton graphs. 
The complete  description of a Calabi-Yau space
  with all its non-trivial $d_i$ fiber structures
needs a full range of n-ary operations where $n_{max}=d+2$.

It has been shown in the toric-geometry approach
how the Dynkin diagrams of affine Cartan-Lie algebras appear in reflexive
K3 polyhedra~\cite{CF,CPR,Greene,KV,Bat}.
We present  an illustratory  example in Fig.(\ref{figxxwe})
where the decomposition  of a K3 polyhedron with 
an elliptic Weierstrass intersection gives as a result two 
Dynkin diagrams for $A_6^{(1)}$ and $E(8)^{(1)}$.
This example is not an isolated one, all the 
elliptic fibrations of K3 spaces found using the UCYA technique feature 
 this kind of decomposition into a pair of graphs that can be 
interpreted as Dynkin diagrams.

\FIGURE[t]{
\parbox{0.6\linewidth}{
\centering
\includegraphics[width=.6\linewidth]{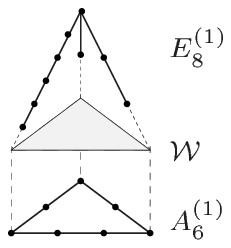}
\label{figxxwe}
\caption{ The schematic decomposition of the the K3 polyhedron determined by 
${\vec k}_4=(1,1,2,3)[7]$ with an elliptic Weierstrass intersection/projection 
(${\cal W}$) 
gives the two  $A_6^{(1)}$ and $E_{(8)}^{(1)}$ Dynkin diagrams.
}}}

As it has been pointed out in the introduction
and as it is obvious from the figure(\ref{figxxwe})
the task of discerning Dynkin diagrams among all the 
set of points
was facilitated by the a priori knowledge of the 
intersecting polyhedra, the Weierstrass triangle in that cases.
UCYA gives naturally the slice structure in the reflexive polyhedra and 
the proyective structure in the corresponding mirror polyhedra.
The knowledge of these slices is a necessary first step in the 
uncovering and understanding of new Dynkin or generalized Dynkin 
diagrams (our Berger diagrams).

It has also been shown~\cite{AENV1,AENV3,AENV4},
using examples of the lattice structure of reflexive polyhedra for CY$_n:  n
\geq 2$ with elliptic fibers,  
 that there is a correspondence
 between the five basic RWVs
(basic constituents of composite RWVs describing  K3 spaces, 
see section 2 in Ref.\cite{Vol}) 
and affine Dynkin diagrams for the five ADE types of Lie
algebras ( A, D  series and exceptional E$_{6,7,8}$).

In each case, a pair of extended RWVs have an
intersection which is a reflexive plane polyhedron; each vector from
the  pair gives the left or right part of the three-dimensional RVW.
  The construction generalizes to any dimension.
In Ref.\cite{Vol}  it was remarked that in the corresponding 
``left'' and right ``graphs'' of 
$CY_{3,4,..}$ Newton reflexive polyhedra one can find 
new graphs with some regularity in its structure.


In principle one should be able to 
 build , classify and understand these regularities 
of the graphs according to the n-arity operation which originated the 
construction.
For the case of binary or arity-2 constructions: two graphs are possible.  
In general for any reflexive polyhedron,  
for a given   arity-r  intersection, it corresponds exactly $r$ graphs.

In the binary case, 
the 2-ary intersection (a plane) in the Newton polyhedra, which 
correspond to the {\em eldest} reflexive vector of the series,
 separate 
left and right graphs.
A concrete rule for the extraction of individual  graph points
from all possible nodes in the graphs 
is that they are selected if they exactly 
 belong ``on the edges''  lying on one side or another with respect the
intersection, see figure(\ref{figxxwe}). 
In the ternary case, 
the  3-ary intersection hypersurface is a volume, which 
 separate three domains in the 
newton polyhedra and three graphs are possible. 
Individual points are assigned to each graph 
looking at their position with respect to 
 the volume intersection (see Tab.(1) in Ref.\cite{Vol}) for 
some aclaratory examples).

The emergence of Dynkin diagrams or generalized Berger diagrams in 
Calabi-Yau reflexive polyhedra is not a mere philatelic 
curiosity: 
in a concrete  singular limit of the K3 space, there appears a gauge 
symmetry whose Cartan-Lie algebra corresponds to the Dynkin diagram seen 
as a graph on one side of Fig.~(\ref{figxxwe}).
In general, the rich singularity structures of K3 $\equiv$ CY$_2$ spaces
are closely connected to the affine Cartan-Lie symmetries A$_r^{(1)}$,
D$_{2r}^{(1)}$, E$_6^{(1)}$, E$_7^{(1)}$ and E$_8^{(1)}$ via the crepant
resolution of specific quotient singular structures such as the
Kleinian-Du-Val singularities ${ {\bf C}^2/G}$~\cite{DuVal}, where $G$ is
a discrete subgroup of SU(2). For example, the crepant resolution of the
${ {\bf C}^2/Z_n}$ singularity gives for rational, i.e., genus-zero, (-2)
curves an intersection matrix that coincides with the A$_{n-1}$ Cartan
matrix. Also, in the case of K3 spaces with elliptic fibers which can be
written in Weierstrass form, there exists and ADE classification of
degenerations of the fibers~\cite{Kodaira,Ber}.


Graphs can directly be obtained from the reflexive polyhedron construction
but can also be defined graphs independently of it.
New graphs will be derived, or by direct manipulation of the original ones,
 or    from 
generalized Cartan matrices in a purely algebraic fashion. They will basically
consist on the primitive graphs extracted from reflexive polyhedra to whom 
internal nodes in the edges will have been added or eliminated.
The nature of the relation, if any, of the graphs thus generated 
to the  geometry of Toric varieties and 
the description of Calabi-Yau as hypersurfaces on them 
is  related to the possibility 
of defining viable ``fan'' lattices. This is an open question, 
clearly 
related  to the properties of the generalized Cartan matrices, interpreted 
as a matrix of divisor intersections.

\section{From Berger graphs to Berger matrices, a algebra review}

Once one has 
 established the existence of Dynkin-like graphs, possibly 
not corresponding to any of the known Lie or affine Kac-Moody algebras,
the next step is to encode the 
 information contained in the graphs  in a more 
workable structure: a matrix of integer numbers to be defined.
If  these  ``Dynkin'' graphs are somehow related 
to possible generalizations of the
 Lie and affine Kac-Moody algebra concepts, 
it is then natural to look for 
possible generalizations of the corresponding affine Kac-Moody Cartan matrices
when searching for possible ways of assigning integral matrices to them.
We include here a little review of some definitions and the procedure of 
the formal definition of Berger matrices already outlined in Ref.\cite{volemi}.
There we mention one  possibility which could serve of guide: 
 to suppose that this affine property remains:
matrices with determinant equal to zero and all principal minors positive.  
We will see in what follows that this is a sensitive choice, on the other 
hand it turn out that the 
usual conditions on the value of the diagonal elements has to be abandoned.

In first place, 
the  building of Cartan-like matrices from already existing graphs is as follows.
We assign to any generalized Dynkin diagram,
a set of  vertices and lines connecting them,
 a matrix, $B$, whose non diagonal elements are 
either zero or are negative integers. 
There are different possibilities, for non diagonal elements,
 considering for the moment
 the most simple case of ``laced'' graphs leading to 
symmetric assignments, we have: 
Case A) there is no line from the vertex $i$ to the vertex $j$. In this 
case the element of the matrix $B_{ij}=0$.
Case B)  there is a single line connecting $i-j$ vertices. In this case 
$B_{ij}=-1$.

The diagonal entries should be defined in addition. 
As a first step, no special restriction is applied and any positive 
integer is allowed. We see however that very quickly  only a few 
possibilities are naturally selected. The diagonal elements of the matrix 
 are two for CY2 originated graphs but are allowed to take increasing 
integer numbers with the dimensionality of the space, 
$3,4...$  for $ CY_{3,4...}$.

A large number of graphs and 
matrices associated to them, obtained by inspection considering 
different possibilities has been checked (see also Ref.\cite{Vol}).  
Some regularities are quickly disclosed.
In first place it is easy to see that there are graphs where
the number of lines outgoing a determined vertex can be bigger than two,
 in  cases of interest
 they will be 3, graphs from CY3, or bigger in the cases of 
graphs coming from CY4 and higher 
dimensional spaces.
Some other important regularities appear. The matrices are genuine generalizations 
of affine matrices. 
Their determinant can be made  equal to zero and all their principal 
minors made positive by careful choice of the diagonal entries depending 
on the Calabi-Yau dimension and n-ary structure. 

Moreover, we can go back to the defining reflexive polyhedra and 
define  other quantities  in purely geometrical terms. For 
example  we can 
consider the position or distance of each of the  vertices of the 
generalized Dynkin diagram to the intersecting reflexive polyhedra.
Indeed, it has been remarked \cite{CF} that Coxeter labels 
for affine Kac-Moody algebras can be obtained 
directly from the graphs: they correspond precisely to this ``distance'' between 
individual nodes and some defined intersection which separates ``left'' and 
``right'' graphs.
Intriguingly, this procedure can be easily generalized to our case, one 
can  see that, by a careful choice of the entry assignment for the 
corresponding  matrix, it follows
 Coxeter labels can be given in a proper way: 
they have the expected 
property of corresponding to the elements of the null vector a generalized 
Cartan matrix.

From the emerging pattern of these  regularities, we are lead to
 define a new set of matrices, generalization 
of Cartan matrices in purely algebraic terms,
the Berger, or Berger-Cartan-Coxeter matrices.
 This will be done in the next paragraph.

Based on previous considerations, we define  now in purely algebraic 
terms \cite{volemi}, the so called Berger Matrices \cite{volemi,Vol}.
We suggest the following rules for them, 
in what follows we will see step by step how they lead
to a consistent construction generalizing the  Affine Kac-Moody concept.
A Berger matrix is a finite integral matrix characterized by the following data:
\begin{eqnarray}
{\mathbb B}_{ii}&=&2,3, 4..\nonumber\\
{\mathbb B}_{ij}& \leq& 0,\quad 
{\mathbb B}_{ij} \in {\mathbb Z} ,\nonumber\\
{\mathbb B}_{ij}=0 &\mapsto & {\mathbb B}_{ji}=0, \nonumber\\
Det\ {\mathbb B} &=&0,\nonumber\\
Det\ {\mathbb B}_{\{(i)\}} &>& 0.\nonumber
\label{eqsberger}
\end{eqnarray}
The last two restrictions, the zero determinant and the positivity of all 
principal proper minors,  corresponds to the  {\it affine condition}. 
They are shared by Kac-Moody Cartan matrices, so we expect that the 
basic definitions and properties of those can be easily generalized.
However, with respect to them, 
we relaxed the restriction on the diagonal elements. Note that,
more than one type of diagonal entry is 
allowed: $2,3,..$ diagonal entries can coexist in a given matrix.

For the sake of convenience, we define also 
{\em ``non-affine'' Berger Matrices}
where the condition of non-zero determinant is again imposed. These matrices 
does not seem to appear naturally resulting from polyhedron graphs 
but they are 
useful when defining root systems 
 for the affine case by extension of them. They 
could play  the same role of basic simple blocks as finite Lie algebras play for the
case of affine Kac-Moody algebras.

The important fact to be remarked here is that this definition 
lead us to a construction with the right properties we would 
expect from a generalization of the Cartan matrix idea.

The systematic enumeration of the various possibilities concerning the 
large family of possible Berger matrices can be facilitated 
by the introduction for each matrix of its generalized Dynkin diagram.
As we intend that the definition of this family of matrices be independent 
of algebraic geometry concepts we need an independent definition of these 
diagrams.  
Obviously the procedure given before 
can be reversed to allow the deduction of the 
generalized Dynkin diagram from its generalized Cartan or Berger Matrix.
An schematic  prescription for the most simple cases could be:
A) For a matrix of dimension $n$, define  $n$ vertices and  draw them
 as small circles. In case of appearance of vertices with different 
diagonal entries, some graphical distinction will be performed.
Consider all the element $i,j$ of the matrix in turn.
B) Draw one line from vertex $i$ to vertex $j$ if the corresponding 
element $A_{ij}$ is non zero.

In what follows, we show that indeed these kind of matrices and 
Dynkin diagrams, exist beyond those  purely defined from 
 Calabi-Yau newton reflexive polyhedra.
In fact we show that there are infinity families of them where 
suggestive regularities appear.

It seems easy to conjecture that 
the set of all, known or generalized,
 Dynkin diagrams obtained from Calabi-Yau spaces 
can be described by this set of Berger matrices. 
It is however not so clear the validity of the opposite   
question,
whether or not the infinite set of generalized Dynkin diagrams 
previously defined can be found digging in the Calabi-Yau $(n,a)$ structure 
indicated by UCYA.
For physical applications however it could be important the following 
remark. Theory of Kac-Moody algebras show us that for any 
{\em finite} or {\em affine} Kac-Moody
 algebra, every proper subdiagram 
(defined as that part of the 
generalized Coxeter-Dynkin diagram obtained by removing one or more vertices 
and the lines attached to these vertices) 
is a collection of diagrams corresponding to {\em finite} Kac-Moody algebras.
In our case we have more flexibility. Proper subdiagrams, obtained 
eliminating internal nodes or vertices, are in general collections of 
Berger-Coxeter-Dynkin diagrams corresponding to other (affine by construction )Berger diagrams 
{\em or} to {\em affine} Kac-Moody algebras.
This property might  open the way to the consideration of non-trivial 
extensions of SM and string symmetries.

Next, one consider the Berger Matrix as a matrix of  inner products in some   root spaces.
Morevoer, for further progress, the interpretation of a Berger matrix as the matrix of 
divisor intersections $B_{ij}\sim D_i\cdot D_j$ in Toric geometry
could be useful for the study of the 
viability of fans of points associated to them, singularity blow-up, and the existence of 
Calabi-Yau varieties itself. This geometrical approach will be pursued 
somewhere else \cite{newtorrente}.
However, for algebraic applications, and with the extension of the CLA and KMA 
concepts in mind, the interpretation of these matrices as matrices 
corresponding to a inner product in some vector space is most natural which 
is our objective now.

The Berger matrices are obtained by weaking the conditions on 
the generalized Cartan matrix ${\hat {\mathbb A}}$ appearing in affine 
Kac-Moody algebras.
In what concern algebraic properties, there are no changes, 
it remains intact the condition of semi-definite positiviness, 
this allows to translate trivially many of 
the basic ideas and terminology for roots and root subspaces 
for  appearing in Kac-Moody 
algebras. Clearly,
the problem of expressing the ``simple'' roots in a orthonormal 
basis was an important step in the classification of semisimple
Cartan-Lie algebras.

For a Berger matrix $ \mathbb B_{ij}$ of dimension $n$, the rank is $r=n-1$.
The  $ (r+1)\times (r+1) $ dimensional is nothing else that a 
generalized Cartan matrix. 
This matrix is symmetric in all the cases of interest in this work.
We expect that a  simple root system
$\Delta^0=\{\alpha_1,\ldots ,\alpha_r \}$ and an extended root
system by $\hat \Delta^0={\alpha_0,\alpha_1,\ldots ,\alpha_r }$,
can be constructed. The defining relation is that the (scaled) 
inner product of the roots is 
\begin{eqnarray}
\alpha_i \cdot \alpha_j &=&\hat {\mathbb B}_{ij}
\qquad 1\leq i,j \leq n.
\end{eqnarray}
The set of roots $\alpha_i$ 
are  the simple roots upon which our generalized Cartan Matrix is based. 
They are supposed to play the analogue of a root basis 
of a semisimple Lie Algebra or of a Kac-Moody algebra.
Note that, as happens in KMA Cartan matrices, 
for having the linearly independent 
set of $\alpha_i$ vectors, we generically  
define them in, at least, a $2n-r$ dimensional space $H$. In our case,
as $r=n-1$, we would need a $n+1$ dimensional space. Therefore, the set 
of $n$ roots satisfying the conditions above has to be completed by 
some additional vector, the ``null root'', to obtain a basis for $H$.
The consideration of these complete set of roots will appear in detail 
elsewhere \cite{newtorrente}.

A generic root, $\alpha$, has the form 
$$\alpha=\sum_i c_i \alpha_i$$
where the set of the coefficients $c_i$ are either all non-negative 
integers or all non-positive integers.
In this $n+1$ dimensional space $H$, generic roots can be defined and the same generalized  definition 
for the inner product of two generic roots $\alpha,\beta$ as in 
affine Kac-Moody algebras applies. This generalized definition 
reduces to the inner product above for any two simple roots.

Since $B$ is of rank $r=n-1$, we can find one, and only one, 
non zero vector $\mu$ such that 
$$B\mu=0.$$
The numbers, $ a_i$, components of the vector $\mu$, 
are called Coxeter labels.
The sums of the Coxeter labels  $h=\sum \mu_i$ is the 
Coxeter number.
For a symmetric generalized Cartan matrix only this type of 
Coxeter number appear.

{\tt to modify?  XXX}

For each  affine matrix we can obtain a number of  non-equivalent 
derived non affine matrix of dimension of smaller dimension
 simply by eliminating one or more of the columns and raws. 
In terms of the graph, this correspond to the elimination of any one of the nodes.
We can explicitly check in all the cases 
that the determinant of these matrices are strictly positive 
and that the matrices are positive definite. 
We can in the same way write the 
set of  roots ${\alpha_i,i=1,\dots,12}$ for this non affine matrix $B^{n-aff}$ such that 
$B_{i,j}^{n-aff}=\alpha_i \cdot \alpha_j$. 
New vectors, fundamental weights, that will play an important role later are
These fundamental weights are defined as the  vectors
 ${\Lambda_i,i=1,\dots,12}$  such that $\delta_{ij}=\Lambda_i\cdot \alpha_j$. 
In the basis of the $\alpha_i's$ they are basically given by the coefficients of the 
inverse of the 
non-affine matrix $B^{n-aff}$.

\section{The  simply laced cases}

Let consider the reflexive polyhedron, which corresponds to a 
K3-sliced $CY_3$
space and which is defined by two extended vectors \cite{Vol}
${\vec k}_L^{ext}, {\vec k}_R^{ext}$. One of these vectors  is coming from 
the set 
$S_L=\{(0,0,0,0,{\vec k }_1),(0,0,0,{\vec k }_2), (0,0,{\vec k }_3),..(perms)..\}$,
where the remaining dots correspond to permutations of the position of 
zeroes and vectors $k$, for example permutations of the 
type $\{ (0,k,0,0,0), (k,0,0,0,0)$, etc \}.
The other defining vector  can  come from the set
$S_R= \{(0,{\vec k }_4), ...(perms)..\} $,  
The vectors ${\vec k}_1,{\vec k}_2,{\vec k}_3$
are respectively any of the five   RWVs of dimension 1,2 and 3. 
The  vector ${\vec k}_4$
correspond to any of the  95 $K3$ RWVs of dimension four. 
As a simple example, a generic quintic CY3 can be 
 defined by two extended vectors,
$\vec k_{1L}^{(ext)}=(1,0,0,0,0)$ 
and $\vec k_{2R}^{(ext)}=(0,1,1,1,1)$ 
(which correspond to the choice $\vec k_4=(1,1,1,1)$). 
The  left and right skeletons
of the reflexive polyhedron  are  determined by extended  vectors,
$\vec k_{1L}^{(ext)},\vec k_{2R}^{(ext)}$ respectively. 
The left skeleton 
will be a tetrahedron with 4-vertices, 6 edges and a number of internal
points over the edges as indicated in the Figure 3 of Ref.\cite{Vol}.

The  RW-simply-laced vectors for dimension 1,2 and 3 and their graphs have already 
been considered before, there are five and only five cases:
\begin{itemize}
\item dim =1 the  vector ${(1)[1]}$ which can be associated to the A series of 
Dynkin diagrams,
\item dim=2, we have the vector ${(11)[2]}$, which is associated to the D series,
\item and dim=3, where the set of vectors  ${(111)[3], (112)[4],(123)[6]}$, 
  correspond, as firstly shown by Candelas and font, to the affine exceptional algebras 
$E_{6,7,8}^{(1)}$.

\end{itemize}

The main objective of this work is to enlarge this list with graphs obtained 
by vectors of dimension four (corresponding to CY3).
In dim=4, corresponding to K3-sliced $CY_3$ spaces, we can single out by inspection the 
 following 14 RW-reflexive vectors from the total of 95- K3-vectors
{(1111),(1122), (1113), (1124), (2334),(1344),(1236),
(1225),(14510),
(1146),(1269),(1,3,8,12),(2,3,10,15)(1,6,14,21)}. 
The graphs corresponding to 
these vectors can easily be obtained as explained before. 
From the geometrical construction Coxeter numbers can be easily assigned to 
each of the nodes of these graphs. 
Moreover we can 
assign genuine Berger matrices to them with specially simple properties: 
they are symmetric, affine (the determinant is zero, the rank one les than the dimension), they lead to the same set of Coxeter as those obtained from the geometrical 
construction. In addition, each of these graphs and matrices seems not 
be ``extendable'': in contradistinction to other cases, see the discussion in 
Ref.\cite{volemi}, no other graphs and Berger matrices can be obtained from them 
simply adding more nodes to any of the legs. In this sense, these graphs are 
``exceptional''. As with the classical exceptional graphs, series can be 
traced among them. Apparently these fourteen vectors are the only ones from the 
the total of 95 vectors which lead to this kind of symmetric matrices.

\subsection{The exceptional simply laced graphs from CY3}

The graphs and matrices
of these simply laced graphs, both, those  already known 
of dimension 1,2,3 and those new of dimension 4 share a number of 
simple characteristics.
The cases of dimension 1,2,3 are well known and correspond to the classical
Cartan Lie algebras. 
Our objective is to give a general description of the new graphs.  
The Berger matrix is obtained from the planar graph according to the standard
rules.  We assign different values (2 or 3 ) to diagonal entries depending 
if they are associated to standard nodes or to the central vertex.
One can assign to  
all of these new graphs a Berger matrix with the following  block structure:
\begin{eqnarray}
B_{SL}&=&
\pmatrix{   
A &0  &0  &0  &v_1 \cr
0 &B  &0  &0  &v_2 \cr
0 &0  &C  &0  &v_3 \cr
0 &0  &0  &D  &v_4 \cr
v_1^t &v_2^t  &v_3^t  &v_4^t  &3 
  }
\end{eqnarray}
where $A,B,C,D$ are square matrices of various dimensions with diagonals filled with two, 
they are the equivalent of the $A_r$ Cartan matrices and the $v_i$ column 
vectors filled with zeroes except for one negative entry, $v_i^t=(0,\dots,0,-1)$.

A generic graph  for anyone of these fourteen vectors is of the form 
depicted in Fig.(\ref{figxxgen}). As we can see in this figure  from a central node four 
legs with respectively $(N_a,N_b,N_c,N_d)$ nodes are attached. Each of the 
legs correspond to one of the regular blocks of the Berger matrix $SL$. The 
central node correspond to the one dimensional block filled with 3 in the matrix.
The Coxeter labels can given in a systematic way \cite{Vol,volemi}, they agree 
with those directly obtained from the matrix $B_{SL}$.
Non affine matrices can be obtained eliminating one or more nodes from the 
legs. Clearly the number of non-affine matrices  depends on the number of 
eliminated nodes and on the symmetry of the diagram. In what follows we will 
list all the non-affine matrices of dimension one less of the original matrix. 
In all these cases can be explicitly checked that the matrices are strictly 
positive definite.

\FIGURE[t]{
\parbox{\linewidth}{
\centering
\includegraphics[width=.5\linewidth]{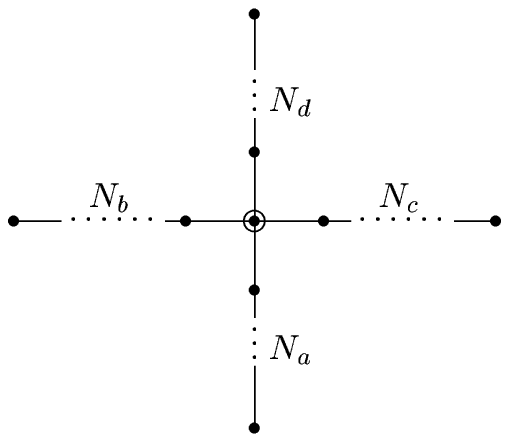}
}
\label{figxxgen}
\caption{The general graph for simply laced cases.}
}

For each diagram, system of roots $\alpha_i$, set of vectors which 
realize  the Berger matrix $SL$ as a matrix of scalar products, 
can be easily obtained. Given a minimal set of orthonormal 
canonical vectors $\{ e_{ai}\}$, one can consider roots of the 
form, for each of the legs 
\begin{eqnarray}
\alpha_{ai}&=& e_{ai}-e_{a,i+1}\nonumber\\
\alpha_{bi}&=& e_{bi}-e_{b,i+1}\nonumber\\
\alpha_{ci}&=& e_{ci}-e_{c,i+1}\nonumber\\
\alpha_{di}&=& e_{di}-e_{d,i+1}\nonumber
\end{eqnarray}
where the set of roots and vectors are assigned for each of the legs 
$l=a,b,c,d$ for the sake of clarity.  
The root corresponding central node, the one corresponding 
to the entry 3 in the matrix, is assigned 
$\alpha_{central}=-\left ( \alpha_{a1}+\alpha_{b1}+\alpha_{c1}+\alpha_{d1}\right )$,
and $\alpha_{l1}$ are roots corresponding to the nearest nodes.
The affine condition is the used to reduce the dimensionality of the space 
spanned by the $e_i's$.
The dimensionality of this space can be furtherly reduced. 
This can be systematically done in a number of ways: we can write  one,two or more
roots as a lineal combination of the rest of them with unknown 
coefficients and ask for the scalar products relations to be fulfilled. 
For the sake of simplicity let us take as a representative
 example any of the cases where 
one of the legs has only one node, i.e. $N_d=1$.
We can write 
\begin{eqnarray}
\alpha_{ai}&=& e_{ai}-e_{a,i+1},\; i=1,\dots,N_a-1 \nonumber\\
\alpha_{bi}&=& e_{bi}-e_{b,i+1},\; i=1,\dots N_b\nonumber\\
\alpha_{ci}&=& e_{ci}-e_{c,i+1},\; i=1,\dots,N_a-1\nonumber\\
\alpha_{cent}&=& -\left ( e_{a1}+e_{b1}+e_{c1}\right ).
\end{eqnarray}
The two roots $\alpha_{a,N_a},\alpha_{c,N_c}$, are obtained by imposing
the scalar products conditions.
\begin{eqnarray}
\alpha_{a,N_a}\cdot \alpha_{a,N_a-1}&=&-1,\quad \alpha_{a,N_a}^2=2,\nonumber \\
\alpha_{c,N_c}\cdot \alpha_{c,N_c-1}&=&-1,\quad \alpha_{c,N_c}^2=2 \nonumber
\end{eqnarray}
 and the mixed relation 
\begin{eqnarray}
\alpha_{a,N_a}\cdot \alpha_{c,N_c}&=&0. \nonumber
\end{eqnarray}
The root corresponding to the fourth leg, $\alpha_{d1}$, is obtained by imposing 
 the affine condition at the end of the procedure. The coefficients 
of the affine condition are the Coxeter labels and these are known from the 
beginning by condition. 
One can check that the following 
expression with arbitrary coefficients $x_j$ satisfy automatically the  
first condition
$\alpha_{a,N_a}\cdot \alpha_{a,N_a-1}=-1$,
\begin{eqnarray}
\alpha_{a,N_a}&=&\frac{1}{x_1+x_2} \left (
x_1 e_{c,N_c}-x_2 \sum_1^{N_c-1} e_{ci}+x_3 \sum_1^{N_b+1} 
e_{bi}+x_4 \sum_1^{N_a} 
e_{ai}
\right )
\nonumber
\end{eqnarray}
and similarly for $\alpha_{c,N_c}$ with arbitrary coefficients $y_j$:
\begin{eqnarray}
\alpha_{c,N_c}&=&\frac{1}{y_1+y_2} \left (
y_1 e_{a,N_a}-y_2 \sum_1^{N_a-1} e_{ai}+y_3 \sum_1^{N_b+1} 
e_{bi}+y_4 \sum_1^{N_c} e_{ci}
\right ).
\nonumber
\end{eqnarray}
These $4+4$ coefficients are constrained  from three non-linear equations 
obtained from the other scalar products.
\begin{eqnarray}
2 \left (x_1+ x_2 \right ) &=& x_1^2 + \left (N_c-1 \right ) x_2^2+\left (N_b+1 \right ) x_3^2+ N_a x_4^2 \nonumber \\
2 \left (y_1+ y_2 \right ) &=&  y_1^2 + \left (N_a-1 \right ) y_2^2+\left (N_b+1 \right ) y_3^2+ N_c y_4^2\nonumber \\
0&=& x_1 y_4+x_4 y_1- \left (N_c-1 \right ) x_2 y_4+\left (N_b+1 \right ) x_3 y_3-\left (N_a-1\right ) x_4 y_2.\nonumber
\end{eqnarray}
Solutions to equations of this type are obtained for a number of cases 
presented on continuation.

We present 
in table~(\ref{t1})
the list of  all the 14 RW vectors of dimension four give above. This table contains all the necessary   information to write down 
the matrices and graphs for each case.
For each of the vectors we give the list of integers $(N_a,N_b,N_c,N_d)$ which 
define both, the number of nodes in each of the four legs of the graph, 
see Fig.(\ref{figxxgen}), and the dimension of the each of the block matrices $A,B,C,D$.
In Figs.(\ref{figall1}-\ref{figall4}) we explicitly give all the graphs with their Coxeter Labels.
On continuation we include in the table, the dimension of the graph or the matrix,
which is equal to the rank plus one and 
 the Coxeter number $h$ which is the sum  the list of the Coxeter labels given in the 
next entry. The last two entries of the table contain information about the 
non-affine derived matrices. The first number is the number of non equivalent 
non-matrices of maximal dimension which can be obtained eliminating one of the nodes 
of the graph (or just one column and row in the respective matrix), the second 
number is the smallest determinant of any of these non-affine derived matrices.
We note that the dimensions of the affine matrices are well bounded in the range
$dim \sim (10,50)$ just above the characteristic dimension of the standard 
affine algebras $E^{(1)}$.
We also note that the total number of non-affine matrices obtained from these 14 
simply laced cases is 34. 
The list of the dimensions of these matrices are $(12,14,16,18,20,25,26,27,28,33,50)$ 
where two series of five and four members can be recognized in addition to two isolated 
dimensions. 
It could be instructive to compare the values of the determinants of these 
non-affine cases with the values for the determinants of the 
Cartan matrices of the well-known non-affine Cartan-Lie algebras
$\det(E_{6,7,8})=3,2,1,\det(F_4,G_2)=1,\det(B_r,C_r)=2,\det D_r=4,\det A_r=r+1$.

\TABLE{
\begin{tabular}{lcccccc}
Vector ${\vec k}_4$& $N_a,N_b,N_c,N_d$ & Dim & h & $(\dots,h_i,dots)$ & $N_{Aff}$ & Det \\ \hline
$(1,1,1,1)[4]$ &(3,3,3,3) & 13&28 &{\small$(3,2,1,3,2,1,3,2,1,3,2,1,4).$}       & 1 & 16  \\
$(2,3,3,4)[12]$   &(2,3,3,5) & 14&90 &{\small$(4,8,3,6,9,3,6,9,2,4,\dots,12 )$}    & 3 &8   \\
$(1,1,1,3)[6]$    &(1,5,5,5) & 17&54 &{\small$(3,1,2,\dots,5,1,2,\dots,5,1,2,\dots,6)$}& 1 &12   \\
$(1,1,2,2)[6]$    &(2,2,5,5) & 15&48 &{\small$(2,4,2,4,1,2,\dots,5,1,2,\dots,6 )$} & 2 &9   \\
$(1,1,2,4)[8]$    &(1,3,7,7) & 19&80 &{\small$(4,2,4,6,1,2,\dots,7,1,2,\dots,8)$}  & 2 & 8   \\
$(1,2,2,5)[10]$   &(1,4,4,9) & 19&100&{\small$(5,2,4,6,8,2,4,6,8,1,2,\dots,10)$}   & 2 &5   \\
$(1,2,3,6)[12]$   &(1,3,5,11)& 21&132&{\small$(6,3,6,9,2,4,\dots,10,1,2,\dots,12)$}& 3 &6   \\
$(1,3,4,4)[12]$   &(2,2,3,11)& 19&120&{\small$(4,8,4,8,3,6,9,1,2,\dots,12)$}       & 3 & 3   \\
$(1,4,5,10)[20]$ &(1,3,4,19)& 28&290&{\small$(10,5,10,15,4,8,\dots,16,1,2,\dots,20)$}& 3&2   \\
$(1,1,4,6)[12]$  &(1,2,11,11)&26 &162&{\small$(6,4,8,1,2,\dots,11,1,2,\dots,12)$}   & 2&6   \\
$(1,2,6,9)[18]$  &(1,2,8,17) &29 &270&{\small$(9,6,12,2,4,\dots,16,1,2,\dots,18)$}  & 3&3   \\
$(1,3,8,12)[24]$&(1,2,7,23) & 34&420&{\small$(12, 8, 16, 3, 6, \dots,21, 1, 2, \dots,24)$}& 3&2   \\
$(2,3,10,15)[30]$&(1,2,9,14)&27&420&{\small $( 5, 10, 20, 3, 6,\dots, 27, 2, 4, \dots,30 )$}& 3&4   \\
$(1,6,14,21)[42]$&(1,2,6,41)&51&1092&{\small$21,14,28,6,12,\dots,36,1,2,\dots,42)$}& 3&1   \\
\hline
\end{tabular}
\label{t1}
\caption{ List of  all the 14 RW vectors of dimension four. The integers $(N_a,N_b,N_c,N_d)$ define both, 
the number of nodes in each of the four legs of the graph
 and the dimension of the each of the block matrices $A,B,C,D$.
The dimension of the matrix ($dim=rank+1$).
 The Coxeter number $h$ and list of Coxeter labels..
The last two entries of correspond to the
non-affine derived matrices. 
First, the number of non equivalent 
non-matrices of maximal dimension which can be obtained eliminating one of the nodes 
of the graph (or just one column and row in the respective matrix), the second 
number is the smallest determinant of any of these non-affine derived matrices.}}


We could ask the question on how we could enlarge this list of  
 affine graphs and matrices  using our Berger construction. 
Following Ref.\cite{volemi}, new matrices could be obtained for example  
starting  from  any of these graph 
 and inserting additional internal nodes. 
However these affine matrices seems to be exceptional, no other affine 
matrices can be obtained from them in this way.

One can also ask the question wether among the graphs and matrices  presented
in table (\ref{t1}) one can find some series in a similar way as the
$E_{6,7,8}$ seem to form a series. Candidates for series like that are the 
graphs of consecutive dimension $(13,14,15)$ and those of the list 
$(26,27,28,29)$. Indeed one can see that the cases of dimension $13,14,15$ present
some similitudes to the $E_{6,7,8}$ series, in particular the root systems 
of the vectors $(1122)$ and $(2334)$ are related to each other in a similar 
way as the $E_7^1,E_8^1$ roots are linked.

In the next paragraphs we will deal in some more detail with each of the fourteen cases in turn, 
paying some more attention to the cases corresponding to the  cases of of lower 
dimension $13,14,15$.

\subsection{Example: the (1111) case.}

We  discuss in some detail the first case apearing in table(2), the 
matrix associated to the vector $(1111)[4]$. 
The Berger matrix is obtained from the planar graph according to the standard
rules.  We assign different values (2 or 3 ) to diagonal entries depending 
if they are associated to standard nodes or to the central vertex.
The result is the following $13\times 13$ symmetric matrix 
containing, as more significant difference, an additional 3 diagonal entry:
{\small
\begin{eqnarray}
CY3B_{()}&=&\left (
\begin{array}{ccc|ccc|ccc|ccc|c}
 2 &-1 & 0   & 0 & 0 & 0   & 0 & 0 & 0   & 0 & 0 & 0  &-1 \\
-1 & 2 &-1   & 0 & 0 & 0   & 0 & 0 & 0   & 0 & 0 & 0  & 0 \\
 0 & -1& 2   & 0 & 0 & 0   & 0 & 0 & 0   & 0 & 0 & 0  & 0 \\
\hline
 0 & 0 & 0  & 2 &-1 & 0    & 0 & 0 & 0   & 0 & 0 & 0  &-1 \\
 0 & 0 & 0  &-1 & 2 &-1    & 0 & 0 & 0   & 0 & 0 & 0  & 0 \\
 0 & 0 & 0  & 0 &-1 & 2    & 0 & 0 & 0   & 0 & 0 & 0  & 0 \\
\hline
 0 & 0 & 0  & 0 & 0 & 0    & 2 &-1 & 0   & 0 & 0 & 0  &-1 \\
 0 & 0 & 0  & 0 & 0 & 0    &-1 & 2 &-1   & 0 & 0 & 0  & 0 \\
 0 & 0 & 0  & 0 & 0 & 0    & 0 & 1 & 2   & 0 & 0 & 0  & 0 \\
\hline
 0 & 0 & 0  & 0 & 0 & 0    & 0 & 0 & 0   & 2 &-1 & 0  &-1 \\
 0 & 0 & 0  & 0 & 0 & 0    & 0 & 0 & 0   &-1 & 2 &-1  & 0 \\
 0 & 0 & 0  & 0 & 0 & 0    & 0 & 0 & 0   & 0 &-1 & 2  & 0 \\
\hline
-1 & 0 & 0  &-1 & 0 & 0    &-1 & 0 & 0   &-1 & 0 & 0  & 3 
\end{array}
\right )
\end{eqnarray}
}
One can check that this matrix fulfills the conditions for 
Berger matrices. Its determinant is zero while the rank $r=12$. 
All the principal minors are positive.

One can obtain a system of roots $(\alpha_i, i=1,\ldots,13 )$ 
in a orthonormal basis.
Considering the  orthonormal canonical basis 
$(\{e_i\}, i=1,\ldots, 12)$, we obtain:
{\small
\begin{eqnarray}
\alpha_1&=& -(e_1-e_2)                                            \nonumber\\
\alpha_2&=& \frac{1}{2} [(e_1 - e_2 - e_3+e_4+e_5+e_6)+(e_8-e_7)]  \nonumber\\
\alpha_3&=&-(e_8- e_7)  \nonumber\\
\alpha_4&=&(e_4- e_3)  \nonumber\\
\alpha_5&=&(e_5- e_4)  \nonumber\\
\alpha_6&=&(e_6- e_5)  \nonumber\\
\alpha_7&=&(e_1+ e_2)  \nonumber\\
\alpha_8&=&   -\frac{1}{2} [(e_1 + e_2 - e_9-e_{10}-e_{11}-e_{12})+(e_8+e_7)]
\nonumber\\
\alpha_9&=& (e_8+e_7)
 \nonumber\\
\alpha_{10}&=& -(e_{10}-e_9) \nonumber\\
\alpha_{11}&=& -(e_{11}-e_{10}) \nonumber\\
\alpha_{12}&=& -(e_{12}-e_{11}) \nonumber\\
\alpha_{13}&=&  e_3-e_2-e_9 \nonumber
\end{eqnarray}
}
The assignment of roots to the nodes of the Berger-Dynkin graph 
is given in Fig.(\ref{fig5}).
It easily to check the inner product of these ¨simple roots¨ 
leads to the Berger Matrix $a_i\cdot a_j=B_{ij}$.
This matrix has one null eigenvector, with coordinates, in the $\alpha$ basis,
$\mu=(3,2,1,3,2,1,3,2,1,3,2,1,4).$
The Coxeter number is $h=22$. One can 
check that these Coxeter labels are identical to those obtained from 
the  geometrical construction \cite{CF,Vol}. They are shown explicitly in 
Fig.(\ref{fig4}).
Correspondingly 
the following linear combination of the ¨roots¨ satisfies the affine 
condition: 
\begin{eqnarray}
4 \alpha_{0}+
3 \alpha_{a1}     + 2 \alpha_{a2}  +     \alpha_{a3}  +
3 \alpha_{b1}     + 2 \alpha_{b2}  +     \alpha_{b3}  +
3 \alpha_{c1}     + 2 \alpha_{c2}  +     \alpha_{c3}  
& =& 0 \nonumber 
 \end{eqnarray}

It is instructive to compare this case with the standard $E_6^{(1)}$ case. 
The graph associated to this case can 
be extracted from a $(111)$ reflexive Newton polyhedron. 
The result appears in Fig.(\ref{fig4},center), we  obtain the 
Coxeter-Dynkin diagram corresponding to the affine algebra $E_6^{(1)}$. 
We can easily check that 
 following the rules given above we can 
form an associated Berger matrix, which, coincides with 
the  corresponding  generalized Cartan matrix of the 
the affine algebra $E_6^{(1)}$. 
The well known Cartan matrix for this is:
\begin{eqnarray}
E_6^{(1)}\equiv CY B3&=&
\pmatrix{ 2&-1&  0& 0&   0& 0&   0\cr
-1& 2&  0& 0&   0& 0&  -1\cr
 0& 0&  2&-1&   0& 0&   0\cr
 0& 0& -1& 2&   0& 0&  -1\cr
 0& 0&  0& 0&   2&-1&   0\cr
 0& 0&  0& 0&  -1& 2&  -1\cr
 0&-1&  0&-1&   0&-1&   2}
\end{eqnarray}
The root system is well known, we have 
(in a, minimal, ortonormal basis $(\{e_i\},i=1,...,8$):
\begin{eqnarray}
\alpha_1   &=& -\frac{1}{2}(-e_1+e_2+e_3+e_4+e_5+e_6+e_7-e_8)\nonumber\\      
\alpha_2   &=&      (e_2-e_1)                             \nonumber\\
\alpha_3   &=&      (e_4-e_3)                             \nonumber\\
\alpha_4   &=&      (e_5-e_4)                             \nonumber\\
\alpha_5   &=&      (e_1+e_2)                             \nonumber\\
\alpha_6   &=& -\frac{1}{2}(e_1+e_2+e_3+e_4+e_5-e_6-e_7+e_8)  \nonumber  \\   
\alpha_7   &=& -e_2+e_3                          \nonumber  
\end{eqnarray}
Coxeter labels and affine condition are easily reobtained.
The diagonalization of the matrix gives us the zero mode vector, 
$B\mu=0$. In this case the Coxeter labels are $\mu=(1,2,1,2,1,2,3)$
 and $h=12$. 
The affine condition satisfied by the 
set of simple roots is also well known 
$\alpha_1+2 \alpha_2+\alpha_3+2\alpha_4+\alpha_5+2 \alpha_6+3\alpha_7=0$.

From the  $(1111)[4]$ $CY3B$ affine matrix we can obtain a derived non affine 
matrix of dimension $12$
 simply by eliminating 
one of the columns and raws. It is straightforward to write the graph for it. 
It is obvious for 
symmetry reasons that in this case there is only one such affine matrix. We can 
explicitly check that the determinant of this matrix is strictly positive 
($det(BE_6)=16$). 
Furthermore
we have checked that the matrix is positive definite. We can in the same way write the 
set of twelve roots ${\alpha_i,i=1,\dots,12}$ for this non affine matrix $B^{n-aff}$ such that 
$B_{i,j}^{n-aff}=\alpha_i \cdot \alpha_j$. 
The fundamental weights 
 ${\Lambda_i,i=1,\dots,12}$  satisfy $\delta_{ij}=\Lambda_i\cdot \alpha_j$. 
In the basis of the $\alpha_i's$ they can be obtained from  the inverse of the 
non-affine matrix $B^{n-aff}$. 
The coefficients of fundamental weights $\Lambda_i$ in this baseis 
are given in the next table:
{\small
\begin{eqnarray}
\begin{array}{c|cc|ccc|ccc|ccc|c}  
 F.W.     &\alpha_{a1}&\alpha_{a2}&\alpha_{b1}&\alpha_{b2}&
\alpha_{b3}&\alpha_{c1}&\alpha_{c2}&
\alpha_{c3}&\alpha_{d1}&\alpha_{d2}&
\alpha_{d3}&\alpha_{0} 
\\\hline
\Lambda_{a1}&6 & 3 & 6    & 4   & 2   & 6    & 4   & 2   & 6    & 4   & 2   & 8 \\
\Lambda_{a2}&3 & 2 & 3    & 2   & 1   & 3    & 2   & 1   & 3    & 2   & 1   & 4 \\\hline
\Lambda_{b1}&6 & 3 & 15/2 & 5   & 5/2 & 27/4 & 9/2 & 9/4 & 27/4 & 9/2 & 9/4 & 9 \\
\Lambda_{b2}&4 & 2 & 5    & 4   & 2   & 9/2  & 3   & 3/2 & 9/2  & 3   & 3/2 & 6 \\
\Lambda_{b3}&2 & 1 & 5/2  & 2   & 3/2 & 9/4  & 3/2 & 3/4 & 9/4  & 3/2 & 3/4 & 3 \\\hline
\Lambda_{c1}&6 & 3 & 27/4 & 9/2 & 9/4 & 15/2 & 5   & 5/2 & 27/4 & 9/2 & 9/4 & 9\\ 
\Lambda_{c2}&4 & 2 & 9/2  & 3   & 3/2 & 5    & 4   & 2   & 9/2  & 3   & 3/2 & 6\\ 
\Lambda_{c3}&2 & 1 & 9/4  & 3/2 & 3/4 & 5/2  & 2   & 3/2 & 9/4  & 3/2 & 3/4 & 3 \\\hline
\Lambda_{d1}&6 & 3 & 27/4 & 9/2 & 9/4 & 27/4 & 9/2 & 9/4 & 15/2 & 5   & 5/2 & 9\\ 
\Lambda_{d2}&4 & 2 & 9/2  & 3   & 3/2 & 9/2  & 3   & 3/2 & 5    & 4   & 2   & 6\\ 
\Lambda_{d3}&2 & 1 & 9/4  & 3/2 & 3/4 & 9/4  & 3/2 & 3/4 & 5/2  & 2   & 3/2 & 3 \\\hline
\Lambda_{0} &8 & 4 & 9    & 6   & 3   & 9    & 6   & 3   & 9    & 6   & 3   & 12\\
\hline
\end{array}
\end{eqnarray}
}

\FIGURE[t]{
\parbox{\linewidth}{\centering
 \unitlength0.7mm
\begin{tabular}{lcr}

   \begin{picture}(50,50)
   \multiput(0,0)(10,0){5}{\circle*{1.7}}
   \put(0,0){\line(1,0){40}}
   \multiput(20,0)(0,10){2}{\circle*{1.7}}
   \put(20,0){\line(0,1){10}}

    \put(1,2){1}
    \put(11,2){2}
    \put(21,2){3}
    \put(31,2){2}
    \put(41,2){1}
    \put(21,2){3}
    \put(21,12){2}
   \end{picture}  
&

   \begin{picture}(50,50)
   \multiput(0,0)(10,0){5}{\circle*{1.7}}
   \put(0,0){\line(1,0){40}}
   \multiput(20,0)(0,10){3}{\circle*{1.7}}
   \put(20,0){\line(0,1){20}}
   
    \put(1,2){1}
    \put(11,2){2}
    \put(21,2){3}
    \put(31,2){2}
    \put(41,2){1}
    \put(21,2){3}
    \put(21,12){2}
    \put(21,22){1}

   \end{picture}

&

\hspace{0.1cm}
 \unitlength0.5mm
   \begin{picture}(50,50)
   \multiput(0,0)(10,0){7}{\circle*{1.7}}
   \put(0,0){\line(1,0){60}}
   \multiput(30,-30)(0,10){7}{\circle*{1.7}}
   \put(30,-30){\line(0,1){60}}
    \put(30,0){\circle{3}}
    \put(1,2){1}
    \put(11,2){2}
    \put(21,2){3}
    \put(31,2){4}
    \put(41,2){3}
    \put(51,2){2}
    \put(61,2){1}
    \put(31,-28){1}
    \put(31,-18){2}
    \put(31,-8){3}
    \put(31,2){4}
    \put(31,12){3}
    \put(31,22){2}
    \put(31,32){1}
   \end{picture}  
\end{tabular}}\vspace{1.4cm}
\label{fig4}
\caption{
Berger-Dynkin diagrams for $E_6$, the 
 affine $E_6^{(1)}$ and its generalization 
$CY3-E_6^{(1)}$ .}
}

One could try to pursue the generalization process of graphs 
and matrices adding internal nodes to this case as it has been done 
previously. Surprisingly, in contradiction to previous case where 
an infinite series of new graphs and matrices can be obtained \cite{volemi}, this is 
however and ``exceptional'' case. No infinite series of graphs can be 
obtained in this way.
Similarly, one can find
  generalizations of 
the $E_7^{[1]}$
and  $E_8^{[1]}$ graphs (corresponding to the choice of three dimensional  
vectors (112),(123)).

\FIGURE[t]{
\parbox{\linewidth}{
\centering
 \unitlength1.2mm
\hspace{-1cm}
   \begin{picture}(50,50)
   \multiput(0,0)(10,0){7}{\circle*{1.7}}
   \put(0,0){\line(1,0){60}}
   \multiput(30,-30)(0,10){7}{\circle*{1.7}}
   \put(30,-30){\line(0,1){60}}
    \put(30,0){\circle{3}}
    \put(0,-5){\small $e_6-e_5$}
    \put(10,-5){\small $e_5-e_4$}
    \put(20,-5){\small $e_4-e_3$}
    \put(30,-5){\small $$}
    \put(40,-5){\small $e_9-e_{10}$ }
    \put(50,-5){\small $e_{10}-e_{11}$ }
    \put(60,-5){\small $e_{11}-e_{12}$ }
    \put(32,-30){\small$e_7+e_8$}
    \put(32,-20){\small$-1/2 (e_2+e_1-e_9-e_{10}-e_{11}-e_{12}+e_7+e_8)$}
    \put(32,-10){\small$e_2+e_1$}
    \put(32,0){\small  $$}
    \put(32,10){\small $e_2-e_1$}
    \put(32,20){\small$-1/2 (e_2-e_1+e_3+e_4+e_5+e_6+e_7-e_8)$}
    \put(32,30){\small $e_7-e_8$}

    \put(15,5){\small $e_3-e_2-e_9$}
    \put(24,5){\vector(1,-1){4}}
   \end{picture}

}
\vspace{4cm}
\label{fig5}
\caption{
Berger-Dynkin diagram and root system for the 
$CY3-E_6^{(1)}$ matrix.}
}


\subsection{Example: the (1122)(6) case  }

In the next example, one 
constructs the Berger  matrix and graph based of the vector ${\vec k}_4= (1122)[6]$ 
from $CY_3$.
The graph associated to this vector  appears in Fig.(\ref{figall1},left). 
The Berger matrix is obtained from the planar graph according to the standard
rules.  We assign different values (2 or 3 ) to diagonal entries depending 
if they are associated to standard nodes or to the central vertex.
The result is the following $15\times 15$ symmetric matrix 
{\small
\begin{eqnarray}
CY3B(1122)&=&\left (
\begin{array}{ccccc|ccccc|cc|cc|c}
 2 &-1 & 0 & 0 & 0    & 0 & 0 & 0 & 0 & 0   & 0 & 0   & 0 & 0     & 0 \\
-1 & 2 &-1 & 0 & 0    & 0 & 0 & 0 & 0 & 0   & 0 & 0   & 0 & 0     & 0 \\
 0 &-1 & 2 & -1 & 0    & 0 & 0 & 0 & 0 & 0   & 0 & 0   & 0 & 0     & 0 \\
 0 & 0 &-1 & 2 &-1    & 0 & 0 & 0 & 0 & 0   & 0 & 0   & 0 & 0     & 0 \\
 0 & 0 & 0 &-1 & 2    & 0 & 0 & 0 & 0 & 0   & 0 & 0   & 0 & 0     &-1 \\
\hline
 0 & 0 & 0 & 0 & 0    & 2 &-1 & 0 & 0 & 0   & 0 & 0   & 0 & 0     & 0 \\    
 0 & 0 & 0 & 0 & 0    &-1 & 2 &-1 & 0 & 0   & 0 & 0   & 0 & 0     & 0 \\ 
 0 & 0 & 0 & 0 & 0    & 0 &-1 & 2 & -1 & 0   & 0 & 0   & 0 & 0     & 0 \\
 0 & 0 & 0 & 0 & 0    & 0 & 0 &-1 & 2 &-1   & 0 & 0   & 0 & 0     & 0 \\ 
 0 & 0 & 0 & 0 & 0    & 0 & 0 & 0 &-1 & 2   & 0 & 0   & 0 & 0     &-1 \\
\hline
 0 & 0 & 0 & 0 & 0    & 0 & 0 & 0 & 0 & 0   & 2 &-1   & 0& 0      & 0 \\
 0 & 0 & 0 & 0 & 0    & 0 & 0 & 0 & 0 & 0   &-1 & 2   & 0& 0      &-1 \\
\hline
 0 & 0 & 0 & 0 & 0    & 0 & 0 & 0 & 0 & 0   & 0 & 0   & 2&-1      & 0 \\
 0 & 0 & 0 & 0 & 0    & 0 & 0 & 0 & 0 & 0   & 0 & 0   &-1& 2      &-1 \\
\hline
-1 & 0 & 0 & 0 & 0    &-1 & 0 & 0 & 0 & 0   &-1 & 0   &-1& 0      & 3 \\
\end{array}
\right )
\end{eqnarray}
}
One can check that this matrix fulfills the conditions for 
Berger matrices. Its determinant is zero while the rank $r=14$.
One can obtain a system of roots $(\alpha_i, i=1,\ldots,15 )$ 
in a orthonormal basis.
Considering the  orthonormal canonical basis 
$(\{e_i\}, i=1,\ldots, 14)$, we obtain:
{\small
\begin{eqnarray}                                  
\alpha_{b5} &=& e_{13}-e_{14}                             \nonumber\\
&\cdots& \nonumber\\
\alpha_{b1} &=& e_{9}-e_{10}                              \nonumber\\
\alpha_{c5} &=& e_{8}-e_{7}                               \nonumber\\
&\cdots& \nonumber\\
\alpha_{c1} &=& e_4-e_3                               \nonumber\\
\alpha_{a2} &=&1/2\left ( e_1-e_2-e_3-e_4-e_5-e_6-e_7-e_8\right )   \nonumber\\
\alpha_{a1} &=& e_2-e_1                               \nonumber\\
\alpha_{d2} &=&-1/2\left ( e_1+e_2-e_9-e_{10}-e_{11}-e_{12}- e_{13}-e_{14}\right ) \nonumber\\
\alpha_{d1} &=& e_2+e_1                               \nonumber\\
\alpha_{0}  &=& e_3-e_2-e_9                            \nonumber
\end{eqnarray}
}
The assignment of roots to the nodes of the Berger-Dynkin graph 
is given according to the notation of  Fig.(\ref{figxxgen}.
It easily to check the inner product of these ¨simple roots¨ 
leads to the Berger Matrix $a_i\cdot a_j=B_{ij}$.
This matrix has one null eigenvector, with coordinates, in the $\alpha$ basis,
$\mu=(2,4,2,4,1,2,...,5,1,2,...,6).$
The Coxeter number is $h=48$. One can 
check that these Coxeter labels are identical to those obtained from 
the  geometrical construction \cite{CF,Vol}. They are shown explicitly in 
Fig.(\ref{figall1},left).
Correspondingly 
the following linear combination of the roots satisfies the affine 
condition:

{\small
\begin{eqnarray}
6 \alpha_{0}+
2 \alpha_{a1}     + 4\alpha_{a2}    +
2 \alpha_{b1}     + 4 \alpha_{b2}    +
1 \alpha_{c1}     + 2 \alpha_{c2}  + \dots +  5 \alpha_{c3}+  
1 \alpha_{d1}     + 2 \alpha_{d2}  + \dots +  5 \alpha_{d3}  
& =& 0 \nonumber 
 \end{eqnarray}
}


\FIGURE[t]{
\parbox{\linewidth}{
\centering
\begin{tabular}{lcr}
\includegraphics[width=.3\linewidth]{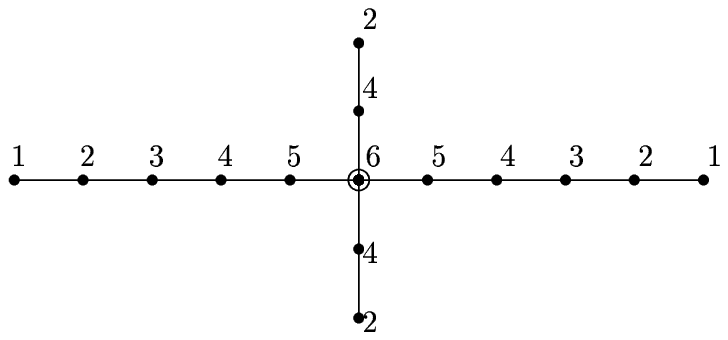} 
&
\hspace{1cm}
\includegraphics[width=.2\linewidth]{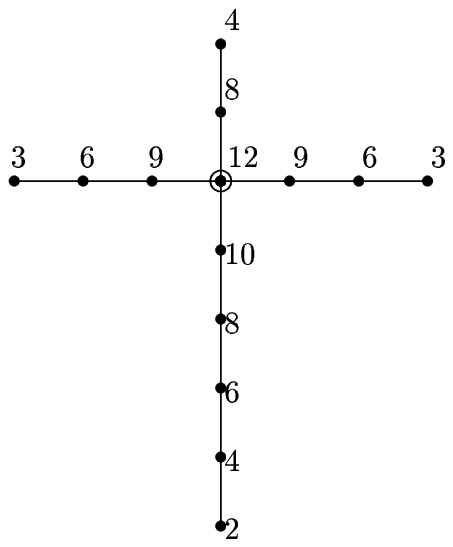} &
\hspace{1cm}
\includegraphics[width=.3\linewidth]{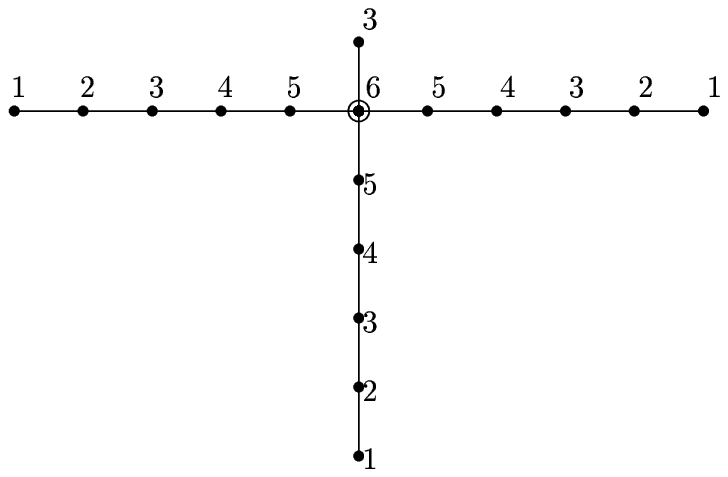}
\end{tabular}
}
\label{figall1}
\caption{Berger Graphs corresponding to the 
vectors (from left to right) 
$(1122)[6]$, $(2334)[12]$ and $(1113)[6]$. 
The Coxeter labels for each node are presented.
}}

For this affine matrix we can obtain {\em two} non-equivalent 
derived non affine matrix of dimension $14$
 simply by eliminating one of the columns and raws. 
In terms of the graph, this correspond to the elimination of any one of the nodes labeled
with Coxeter labels $1$ or $2$.
We can explicitly check that the determinant of these matrices are strictly positive 
($det=10,32$ for elimination Coxeter label nodes $1,2$ respectively). 
Furthermore
we have checked that the matrices are positive definite. 
We can in the same way write the 
set of  roots ${\alpha_i,i=1,\dots,12}$ for this non affine matrix $B^{n-aff}$ such that 
$B_{i,j}^{n-aff}=\alpha_i \cdot \alpha_j$. 
The fundamental weights are defined as before. The ones
 corresponding to the elimination of the 
Coxeter label 1 node are given in table(\ref{tw1122}).

\TABLE{
\label{tw1122}
\caption{Coefficients of the fundamental weights $\Lambda_i$
with respect the $\alpha_i$ basis. 
Non affine matrix obtained from the elimination of 
the root with Coxeter label 1 from the vector $(1122)[6]$.}
\parbox{\linewidth}{
{\small
\begin{eqnarray}
\begin{array}{c|cccc|ccccc|cc|cc|c}
& \alpha_{b4}&\alpha_{b3}&\alpha_{b2}&\alpha_{b1}&\alpha_{c5}&\alpha_{c4}&
\alpha_{c3}&\alpha_{c2}&\alpha_{c1}&\alpha_{a3}&\alpha_{a2}&\alpha_{d3}&\alpha_{d2}&\alpha_{0}\\ \hline
\Lambda_{b4}& 2&3&4&5&1&2&3&4&5&2&4&2&4&6\\
\Lambda_{b3}&  3&6&8&10&2&4&6&8&10&4&8&4&8&12 \\
\Lambda_{b2}&  4&8&12&15&3&6&9&12&15&6&12&6&12&18 \\
\Lambda_{b1}& 5&10&15&20&4&8&12&16&20&8&16&8&16&24 \\ \hline
\Lambda_{c5}& 1&2&3&4&5/3&7/3&3&11/3&13/3&5/3&10/3&5/3&10/3&5 \\
\Lambda_{c4}& 2&4&6&8&7/3&14/3&6&22/3&26/3&10/3&20/3&10/3&20/3&10 \\
\Lambda_{c3}& 3&6&9&12&3&6&9&11&13&5&10&5&10&15 \\
\Lambda_{c2}& 4&8&12&16&11/3&22/3&11&44/3&52/3&20/3&40/3&20/3&40/3&20 \\
\Lambda_{c1}& 5&10&15&20&13/3&26/3&13&52/3&65/3&25/3&50/3&25/3&50/3&25\\ \hline
\Lambda_{a3}& 2&4&6&8&5/3&10/3&5&20/3&25/3&4&7&10/3&20/3&10 \\
\Lambda_{a2}& 4&8&12&16&10/3&20/3&10&40/3&50/3&7&14&20/3&40/3&20 \\ \hline
\Lambda_{d3}& 2&4&6&8&5/3&10/3&5&20/3&25/3&4&7&10/3&20/3&10 \\
\Lambda_{d2}& 4&8&12&16&10/3&20/3&10&40/3&50/3&7&14&20/3&40/3&20 \\
\Lambda_{0}& 6&12&18&24&5&10&15&20&25&10&20&10&20&30 \\ \hline
\end{array}\nonumber
\end{eqnarray}
}}}

\subsection{Example: the $(2334)$ and the rest of SL  
Berger graphs  }

\FIGURE[t]{
\label{figall2}
\caption{
Berger Graphs corresponding to the 
vectors (from left to right) 
 $(1124)[8]$ ,    $(1225)[10]$ and $(1236)[12]$.}  
\parbox{\linewidth}{
\centering
\begin{tabular}{lcr}
\includegraphics[width=.3\linewidth]{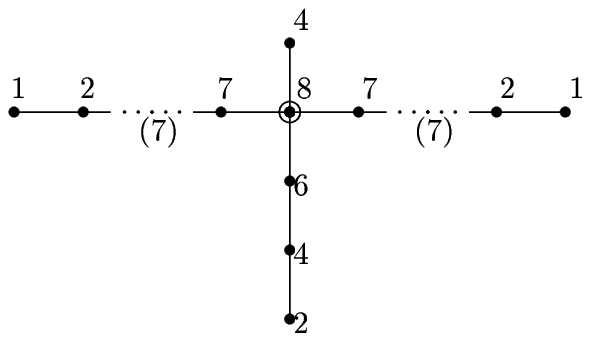} &
\hspace{0.5cm}
\includegraphics[width=.3\linewidth]{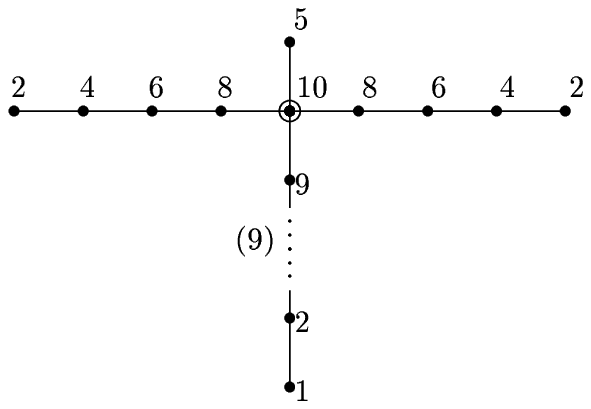} &
\hspace{0.5cm}
\includegraphics[width=.3\linewidth]{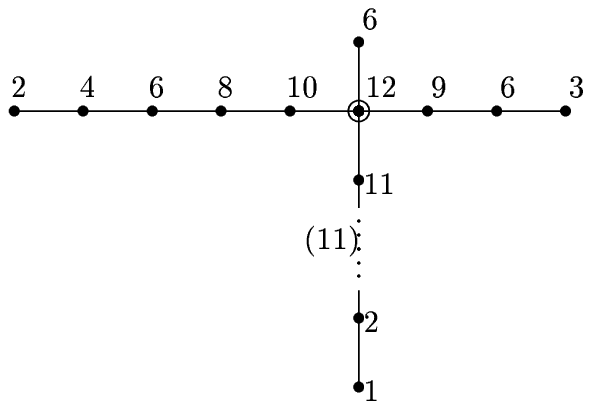} 
\end{tabular}
}}

\FIGURE[t]{
\label{figall3}
\caption{Berger Graphs corresponding to the 
vectors (from left to right) 
$(1344)[12]$,   $(145,10)[20]$  and $(1146)[12]$.}  
\parbox{\linewidth}{
\centering
\begin{tabular}{lcr}
\includegraphics[width=.22\linewidth]{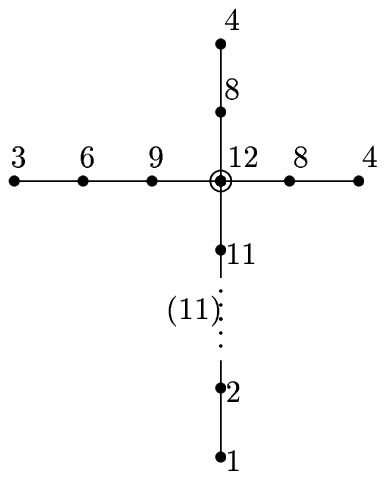} &
\hspace{1cm}
\includegraphics[width=.3\linewidth]{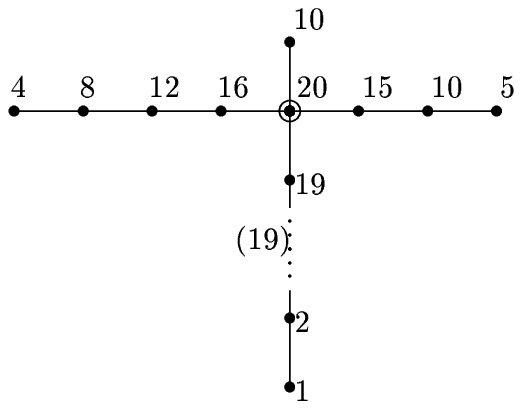} &
\hspace{1cm}
\includegraphics[width=.27\linewidth]{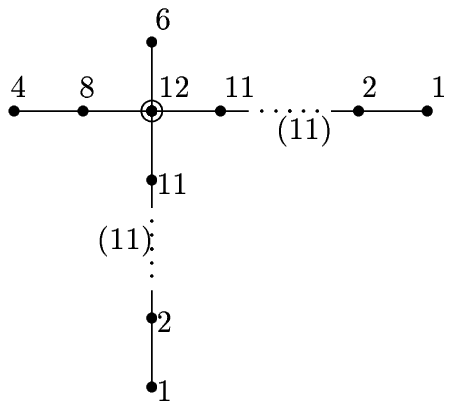} 
\end{tabular}
}}

\FIGURE[t]{
\label{figall4}
\caption{Berger Graphs corresponding to the 
vectors (from left to right) 
$(1269)[18]$,  $(138,12)[24]$,
 $(23,10,15)[30]$ and $(16,14,21)[42]$.}
\parbox{\linewidth}{
\centering
\begin{tabular}{lr}
\includegraphics[width=.3\linewidth]{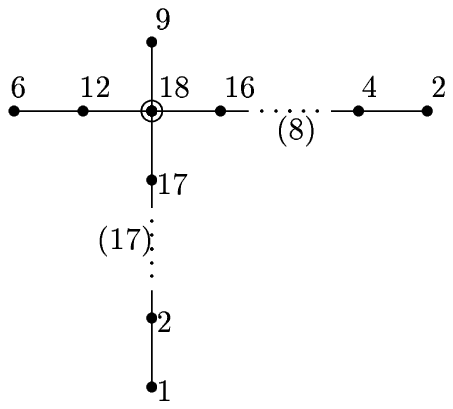} &
\hspace{2cm}
\includegraphics[width=.3\linewidth]{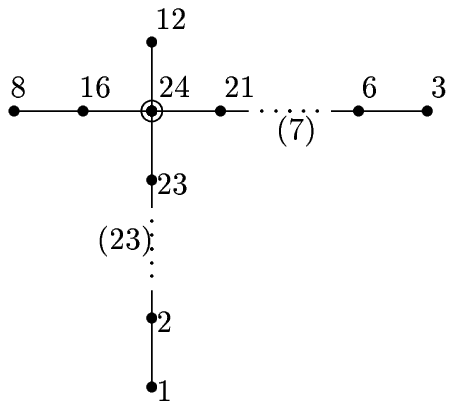}\\[1cm]
\includegraphics[width=.3\linewidth]{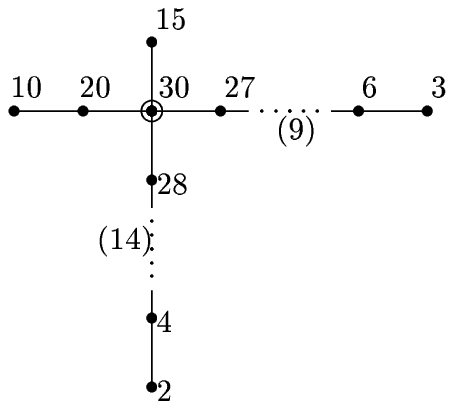} &
\hspace{1cm}
\includegraphics[width=.3\linewidth]{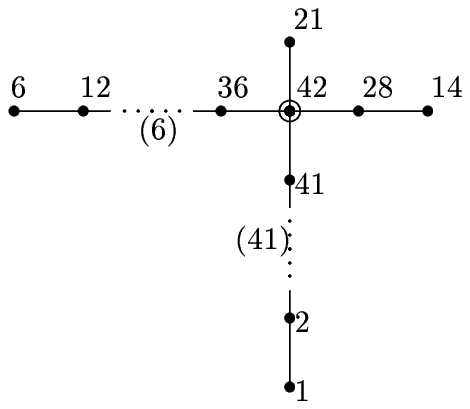} 
\end{tabular}
}}

As before, 
one can construct the Berger  matrix and graph based of the vector 
${\vec k}_4= (2334)[12]$ from $CY_3$.
The graph associated to this vector  appears in Fig.(\ref{figall1},center). 
The result is the following, rank $=14$, $15\times 15$ symmetric matrix 
{\small
\begin{eqnarray}
CY3B(2334)&=&\left (
\begin{array}{cccccc|ccc|ccc|cc|c}
 2 &-1 & 0 & 0 & 0 & 0   & 0 & 0 & 0   & 0 & 0 & 0   & 0 & 0   & 0 \\
-1 & 2 &-1 & 0 & 0 & 0   & 0 & 0 & 0   & 0 & 0 & 0   & 0 & 0   & 0 \\
 0 &-1 & 2 & 0 & 0 & 0   & 0 & 0 & 0   & 0 & 0 & 0   & 0 & 0   & 0 \\
 0 & 0 &-1 & 2 &-1 & 0   & 0 & 0 & 0   & 0 & 0 & 0   & 0 & 0   & 0 \\
 0 & 0 & 0 &-1 & 2 &-1   & 0 & 0 & 0   & 0 & 0 & 0   & 0 & 0   & 0 \\
 0 & 0 & 0 & 0 &-1 & 2   & 0 & 0 & 0   & 0 & 0 & 0   & 0 & 0   &-1 \\
\hline
 0 & 0 & 0 & 0 & 0 & 0   & 2 &-1 & 0   & 0 & 0 & 0   & 0 & 0   & 0 \\    
 0 & 0 & 0 & 0 & 0 & 0   &-1 & 2 &-1   & 0 & 0 & 0   & 0 & 0   & 0 \\ 
 0 & 0 & 0 & 0 & 0 & 0   & 0 &-1 & 2   & 0 & 0 & 0   & 0 & 0   &-1 \\
\hline
 0 & 0 & 0 & 0 & 0 & 0   & 0 & 0 & 0   & 2 &-1 & 0   & 0 & 0   & 0 \\ 
 0 & 0 & 0 & 0 & 0 & 0   & 0 & 0 & 0   &-1 & 2 &-1   & 0 & 0   & 0 \\
 0 & 0 & 0 & 0 & 0 & 0   & 0 & 0 & 0   & 0 &-1 & 2   & 0 & 0   &-1 \\
\hline
 0 & 0 & 0 & 0 & 0 & 0   & 0 & 0 & 0   & 0 & 0 & 0   & 2 &-1   & 0 \\
 0 & 0 & 0 & 0 & 0 & 0   & 0 & 0 & 0   & 0 & 0 & 0   &-1 & 2   &-1 \\
\hline
 0 & 0 & 0 & 0 & 0 &-1   & 0 & 0 &-1   & 0 & 0 &-1   & 0 &-1   & 3 \\
\end{array}
\right )
\end{eqnarray}
}
A system of roots $(\alpha_i, i=1,\ldots,15 )$ 
in a orthonormal basis.
Considering the  orthonormal canonical basis 
$(\{e_i\}, i=1,\ldots, 14)$, we obtain:
{\small
\begin{eqnarray}                                    \nonumber\\
\alpha_{a5} &=& e_{13}-e_{14}                             \nonumber\\
&\cdots& \nonumber \\
\alpha_{a1} &=& e_{9}-e_{10}                              \nonumber\\
\alpha_{b3} &=& e_7+e_8                               \nonumber\\
\alpha_{b2} &=&-1/2 \left ( e_2-e_1+ e_3+e_4+e_5+e_6+ e_7+e_8\right )      \nonumber\\
\alpha_{b1} &=& e_2-e_1                               \nonumber\\
\alpha_{c3} &=& e_6-e_5                               \nonumber\\
\alpha_{c2} &=& e_5-e_4                               \nonumber\\
\alpha_{c1} &=& e_4-e_3                               \nonumber\\
\alpha_{d2} &=&-1/2\left (e_1+e_2-e_9-e_{10}-e_{11}-e_{12}-e_{13}-e_{14}\right )   \nonumber\\
\alpha_{d1} &=& e_1+e_2                               \nonumber\\
\alpha_{0}  &=& e_3-e_2-e_9                            \nonumber
\end{eqnarray}
}
The  null eigenvector has  coordinates in the $\alpha$ basis, or 
Coxeter labels,
$\mu=(4,8,3,6,9,3,6,9,2,4,6,8,10,12).$
The Coxeter number is $h=90$.
For this affine matrix we can obtain {\em three} non-equivalent 
derived non affine, positive definite, matrices of dimension $14$
 simply by eliminating one of the columns and raws. 
They correspond to the elimination of any one of the extreme  nodes labeled
with Coxeter labels $2,3,4$. The determinants are $8,18,32$.
The fundamental weights corresponding to the elimination of the 
Coxeter label 2,3 and 4  nodes are given in 
tables (\ref{tw2334a},\ref{tw2334b},\ref{tw2334c}) respectively.

The analysis of the rest of the graphs, matrices and obtention of roots and 
vectors is completely similar to the examples presented until now and offer no 
difficulty: all the information neccesary to recover these cases have been already 
presented in table (\ref{t1}). The complete list of the graphs is explictly 
presented in the Figs.(\ref{figall1},\ref{figall2},\ref{figall3},\ref{figall4}).
Additional examples of root systems are presented in 
tables(\ref{troots1},\ref{troots2})
and those of weight vectors in 
table(\ref{tw1113}).

\TABLE{
\label{tw2334a}
\caption{Coefficients of the fundamental weights $\Lambda_i$
with respect the $\alpha_i$ basis. 
Non affine matrix obtained from the elimination of 
the root with Coxeter label 2 from the vector $(2334)[12]$.}
{\small
\begin{tabular}{c|cccc|ccc|ccc|cc|c}
      &$\alpha_{a4}$&$\alpha_{a3}$&$\alpha_{a2}$&$\alpha_{a1}$&
$\alpha_{b3}$&$\alpha_{b2}$&$\alpha_{b1}$&
$\alpha_{c3}$&$\alpha_{c2}$&$\alpha_{c1}$&
$\alpha_{d2}$&$\alpha_{d1}$&$\alpha_{0}$ \\ \hline
$\Lambda_{a4}$&2   &  3 &  4  &  5  & 3/2 & 3   & 9/2 &3/2 & 3  & 9/ 2& 2  & 4  &  6   \\
$\Lambda_{a3}$&3   &  6 &  8  & 10  &  3  & 6   & 9   & 3  & 6  & 9   & 4  & 8  & 12   \\
$\Lambda_{a2}$&4   &  8 & 12  & 15  & 9/2 & 9   & 27/2&9/2 & 9  &27/2 & 6  & 12 & 18   \\
$\Lambda_{a1}$&5   & 10 & 15  & 20  &  6  &12   & 18  & 6  & 12 & 18  & 8  & 16 & 24   \\ 
\hline
$\Lambda_{b3}$&3/2 &  3 & 9/2 &  6  &21/8 &17/4 &47/8 &15/8&15/4&45/8 &5/2 & 5  &15/2  \\
$\Lambda_{b2}$&3   &  6 & 9   & 12  &17/4 &17/2 &47/4 &15/4&15/2&45/4 & 5  & 10 & 15   \\  
$\Lambda_{b1}$&9/2 &  9 & 27/2& 18  &47/8 &47/4 &141/8&45/ &45/4&135/8&15/2& 15 & 45/2 \\  
\hline
$\Lambda_{c3}$&3/2 &  3 & 9/2 &  6  & 15/8&15/4 &45/8 &21/8&17/4&47/8 &5/2 & 5  & 15/2 \\ 
$\Lambda_{c2}$&3   &  6 &  9  & 12  &15/4 &15/2 &45/4 &17/4&17/2& 47/4&  5 &10  & 15   \\  
$\Lambda_{c1}$&9/2 &  9 & 27/2& 18  &45/8 &45/4 &135/8&47/8&47/4& 141/8&15/2&15 & 45/2 \\ 
\hline
$\Lambda_{d2}$&2   &  4 & 6   & 8   & 5/2 &  5  &15/2 &5/2 &  5 & 15/2 & 4  & 7 & 10   \\ 
$\Lambda_{d1}$&4   &  8 & 12  & 16  &  5  & 10  &15   &  5 & 10 & 15  &  7  & 14& 20   \\ 
\hline
$\Lambda_{0}$&6   &12  &  18 & 24  & 15/2&  15 &45/2 &15/2& 15 &45/2 & 10  & 20& 30\\  \hline
\end{tabular}
}}

\TABLE{
\label{tw2334b}
\caption{Coefficients of the fundamental weights $\Lambda_i$
with respect the $\alpha_i$ basis. 
Non affine matrix obtained from the elimination of 
the root with Coxeter label 3 from the vector $(2334)[12]$.}
\parbox{\linewidth}{
{\small
\begin{eqnarray}
\begin{array}{c|ccccc|cc|ccc|cc|c}
 F.W.     &\alpha_{a5}&\alpha_{a4}&\alpha_{a3}&\alpha_{a2}&\alpha_{a1}&
\alpha_{b2}&\alpha_{b1}&
\alpha_{c3}&\alpha_{c2}&\alpha_{c1}&
\alpha_{d2}&\alpha_{d1}&\alpha_{0} 
\\\hline
\Lambda_{a5}&7/6  &  4/3 &  3/2 &  5/3 &  11/6 & 2/3 &  4/3 &  1/2 &  1 &  3/2 & 2/3 & 
4/3  &  2  \\
\Lambda_{a4}&4/3  &  8/3 &  3   & 10/3 &  11/3 & 4/3 &  8/3 &  1   &  2 &  3   & 4/3 &  
8/3  &  4  \\
\Lambda_{a3}&3/2  &  3   &  9/2 &  5   &  11/2 & 2   &  4   &  3/2 &  3 &  9/2 & 2   &  
4    &  6  \\
\Lambda_{a2}&5/3  &  10/3&  5   & 20/3 &  22/3 & 8/3 & 16/3 &  2   &  4 &  6   & 8/3 & 
16/3 &  8 \\
\Lambda_{a1}&11/6 &  11/3& 11/2 & 22/3 &  55/6 & 10/3& 20/3 &  5/2 &  5 & 15/2 &10/3 &
20/3 &  10  \\\hline
\Lambda_{b2}&2/3  &  4/3 &  2   & 8/3  &  10/3 &  2  &  3   &  1   &  2 &  3   &4/3  & 
8/3  &  4   \\
\Lambda_{b1}&4/3  &  8/3 &  4   & 16/3 &  20/3 &  3  &  6   &  2   &  4 &  6   & 8/3 &  
16/3 &  8   \\\hline
\Lambda_{c3}&1/2  &  1   &  3/2 & 2    &  5/2  &  1  &  2   & 3/2  &  2 &  5/2 & 1   &  
2    &  3   \\
\Lambda_{c2}&1    &  2   &  3   &  4   &  5    &  2  &  4   &  2   &  4 & 5    &  2  &  
4    &  6   \\
\Lambda_{c1}&3/2  &  3   & 9/2  &  6   &  15/2 &  3  &  6   &  5/2 &  5 &  15/2&  3  &  
6    &  9   \\\hline
\Lambda_{d2}&2/3  &  4/3 &  2   &  8/3 & 10/3  &  4/3&  8/3 &  1   &  2 &  3   &  2  &  
3    &  4   \\
\Lambda_{d1}&4/3  &  8/3 &  4   & 16/3 & 20/3  &  8/3& 16/3 &  2   &  4 &  6   & 3   &   
6    &   8  \\ \hline
\Lambda_{0}&2    &  4   &  6   &  8   & 10    &  4  &  8   &  3   &  6 &  9   & 4   &  
8    &  12  \\ \hline
\end{array}
\end{eqnarray}
}}}

\TABLE{
\label{tw2334c}
\caption{Coefficients of the fundamental weights $\Lambda_i$
with respect the $\alpha_i$ basis. 
Non affine matrix obtained from the elimination of 
the root with Coxeter label 4 from the vector $(2334)[12]$.}
\parbox{\linewidth}{
{\small
\begin{eqnarray}
T4=
\begin{array}{c|ccccc|ccc|ccc|c|c}
 F.W.     &\alpha_{a5}&\alpha_{a4}&\alpha_{a3}&\alpha_{a2}&\alpha_{a1}&
\alpha_{b3}&\alpha_{b2}&\alpha_{b1}&
\alpha_{c3}&\alpha_{c2}&\alpha_{c1}&
\alpha_{d1}&\alpha_{0} 
\\\hline
\Lambda_{a5}&1    &  1   &  1  & 1  &  1    & 1/4  & 1/2  & 3/4  & 1/4  & 1/2  & 
3/4  &  1/2 &  1 \\
\Lambda_{a4}&1    &  2   &  2  & 2  &  2    & 1/2  & 1    & 3/2  & 1/2  & 1    &
3/2  &  1   &  2 \\
\Lambda_{a3}&1    &  2   & 3   & 3  &  3    & 3/4  & 3/2  & 9/4  & 3/4  & 3/2  &
9/4  &  3/2 & 3 \\
\Lambda_{a2}&1    &  2   & 3   & 4  &  4    & 1    & 2    & 3    & 1    & 2    & 
3    &  2   & 4 \\
\Lambda_{a1}&1    &  2   & 3   & 4  &  5    & 5/4  & 5/2  & 15/4 & 5/4  & 5/2  &
15/4 &  5/2 & 5 \\\hline
\Lambda_{b3}&1/4  &  1/2 & 3/4 & 1  &  5/4  & 9/8  & 5/4  & 11/8 & 3/8  & 3/4  & 
9/8  &  3/4 & 3/2 \\
\Lambda_{b2}&1/2  &  1   & 3/2 & 2  &  5/2  & 5/4  & 5/2  & 11/4 & 3/4  & 3/2  &
9/4  &  3/2 & 3  \\
\Lambda_{b1}&3/4  &  3/2 & 9/4 & 3  &  15/4 & 11/8 & 11/4 & 33/8 & 9/8  & 9/4  &
27/8 &  9/4 & 9/2\\\hline
\Lambda_{c3}&1/4  &  1/2 & 3/4 & 1  &  5/4  & 3/8  & 3/4  & 9/8  & 9/8  & 5/4  &
11/8 &  3/4 & 3/2 \\
\Lambda_{c2}&1/2  &  1   & 3/2 & 2  &  5/2  & 3/4  & 3/2  & 9/4  & 5/4  & 5/2  & 
11/4 &  3/2 & 3 \\
\Lambda_{c1}&3/4  &  3/2 & 9/4 & 3  &  15/4 & 9/8  & 9/4  & 27/8 & 11/8 & 11/4 &
33/8 &  9/4 & 9/2\\\hline
\Lambda_{d1}&1/2  &  1   & 3/2 & 2  &  5/2  & 3/4  & 3/2  & 9/4  & 3/4  & 3/2  & 
9/4  &  2   & 3 \\\hline
\Lambda_{0}&1    &  2   & 3   & 4  &  5    & 3/2  & 3    & 9/2  & 3/2  & 3    & 
9/2  &  3   & 6\\\hline
\end{array}
\end{eqnarray}
}}}

\TABLE{
\label{tw1113}
\caption{Coefficients of the fundamental weights $\Lambda_i$
with respect the $\alpha_i$ basis. 
Non affine matrix obtained from the elimination of 
one of the root   from the vector $(1113)[6]$.}
\parbox{\linewidth}{
{\small
\begin{eqnarray}
\begin{array}{c|cccc|ccccc|ccccc|c|c}
 F.W.   &\alpha_{a4}&\alpha_{a3}&\alpha_{a2}&\alpha_{a1}&
\alpha_{b5}&\alpha_{b4}&\alpha_{b3}&\alpha_{b2}&\alpha_{b1}&
\alpha_{c5}&\alpha_{c4}&\alpha_{c3}&\alpha_{c2}&\alpha_{c1}&
\alpha_{d1}&\alpha_{0} 
\\\hline
\Lambda_{a4}&2 & 3 & 4 & 5 & 1 & 2 & 3 & 4 & 5 & 1 & 2 & 3 & 4 & 5 & 3 & 6 \\
\Lambda_{a3}&3 & 6 & 8 & 10 &  2 & 4 & 6 & 8 & 10 & 2 & 4 & 6 & 8 & 10 & 6 & 12 \\
\Lambda_{a2}&4 & 8 & 12 & 15 & 3 & 6 & 9 & 12 & 15 & 3 & 6 & 9 & 12 & 15 & 9 & 18\\
\Lambda_{a1}&5 & 10 & 15 & 20 & 4 & 8 & 12 & 16 & 20 & 4 & 8 & 12 & 16 & 20 & 12 & 24\\
 \hline
\Lambda_{b5}&1 & 2 & 3 & 4 & 5/3 & 7/3 & 3 & 11/3 & 13/3 & 5/6 & 5/3 & 5/2 & 10/3 & 25/6 & 5/2 & 5\\
\Lambda_{b4}&2 & 4 & 6 & 8 & 7/3 & 14/3 & 6 & 22/3 & 26/3 & 5/3 & 10/3 & 5 & 20/3 & 25/3 & 5 & 10\\
\Lambda_{b3}&3 & 6 & 9 & 12 & 3 & 6 & 9 & 11 & 13 & 5/2 & 5 & 15/2 & 10 & 25/2 & 15/2 & 15\\
\Lambda_{b2}&4 & 8 & 12 & 16 & 11/3 & 22/3 & 11 & 44/3 & 52/3 & 10/3 & 20/3 & 10 & 40/3 & 50/3 & 10 & 20 
 \\ \hline
\Lambda_{b1}&5 & 10 & 15 & 20 & 13/3 & 26/3 & 13 & 52/3 & 65/3 & 25/6 & 25/3 & 25/2 & 50/3 & 125/6 & 25/2 & 25\\
\Lambda_{c5}&1 & 2 & 3 & 4 & 5/6 & 5/3 & 5/2 & 10/3 & 25/6 & 5/3 & 7/3 & 3 & 11/3 & 13/3 & 5/2 & 5\\
\Lambda_{c4}&2 & 4 & 6 & 8 & 5/3 & 10/3 & 5 & 20/3 & 25/3 & 7/3 & 14/3 & 6 & 22/3 & 26/3 & 5 & 10\\
\Lambda_{c3}&3 & 6 & 9 & 12 & 5/2 & 5 & 15/2 & 10 & 25/2 & 3 & 6 & 9 & 11 & 13 & 15/2 & 15  \\
\Lambda_{c2}4 & 8 & 12 & 16 & 10/3 & 20/3 & 10 & 40/3 & 50/3 & 11/3 & 22/3 & 11 & 44/3 & 52/3 & 10 & 20 
 \\
\Lambda_{c1}&5 & 10 & 15 & 20 & 25/6 & 25/3 & 25/2 & 50/3 & 125/6 & 13/3 & 26/3 & 13 & 52/3 & 65/3 & 
25/2 & 25\\ \hline
\Lambda_{d1}&3 & 6 & 9 & 12 & 5/2 & 5 & 15/2 & 10 & 25/2 & 5/2 & 5 & 15/2 & 10 & 25/2 & 8 & 15\\
 \hline
\Lambda_{0}&6 & 12 & 18 & 24 & 5 & 10 & 15 & 20 & 25 & 5 & 10 & 15 & 20 & 25 & 15 & 30 \\
\hline
\end{array}
\end{eqnarray}
}}}

\TABLE{
\label{troots1}
\caption{ 
Roots for the Berger cases (Left) $(1236)[12]$ and
(Right)  $(1344)[12]$ where $p=\sqrt{3}$.}
\begin{tabular}{lr}
\parbox{0.49   \linewidth}{
{\scriptsize
\begin{eqnarray}                                
\alpha_{a11}&=& e_{19}-e_{20}                             \nonumber\\
  &\dots & \nonumber \\
\alpha_{a2} &=& e_{10}-e_{11}                             \nonumber\\
\alpha_{a1} &=& e_9-e_{10}                              \nonumber\\
\alpha_{b5} &=&-1/2\left ( e_7-e_6-e_5-e_4-e_3-e_2-e_1-e_8\right )        \nonumber\\
\alpha_{b4} &=& e_7-e_6                               \nonumber\\
&\cdots&\nonumber\\
\alpha_{b1} &=& e_4-e_3                               \nonumber\\
\alpha_{c3} &=& 1/3\left ( 2e_8-e_1-e_2+e_9+e_{10}+...+e_{20}\right )       \nonumber\\
\alpha_{c2} &=& e_1-e_8                               \nonumber\\
\alpha_{c1} &=& e_2-e_1                               \nonumber\\
\alpha_{d1} &=& 1/2\left (e_2+e_1+e_8-e_3-e_4-e_5-e_6-e_7\right )        \nonumber\\
\alpha_{0}  &=& e_3-e_2-e_9                            \nonumber
\end{eqnarray}
}}
&
\parbox{0.50\linewidth}{
{\scriptsize
\begin{eqnarray}                                
\alpha_{a11}&=& e_{17}-e_{18}                             \nonumber\\
&\cdots & \nonumber \\
\alpha_{a1} &=& e_7-e_8                               \nonumber\\
\alpha_{b3} &=& 1/3\left  (2 e_5-e_1+e_7+e_8+...+e_{18}\right )           \nonumber\\
\alpha_{b2} &=& e_1-e_5                               \nonumber\\
\alpha_{b1} &=& e_2-e_1                               \nonumber\\
\alpha_{c2} &=& -1/2\left ( e_4-e_3-e_2-e_1-e_5+p e_6\right )            \nonumber\\
\alpha_{c1} &=& e_4-e_3                               \nonumber\\
\alpha_{d2} &=& -1/2\left (e_4+e_3+e_2+e_1+e_5+p e_6\right )            \nonumber\\
\alpha_{d1} &=& e_4+e_3                               \nonumber\\
\alpha_{0}  &=& e_3-e_2-e_7                            \nonumber
\end{eqnarray}
}
}
\end{tabular}
}

\TABLE{
\label{troots2}
\caption{ Roots for the Berger cases (Left) $(1146)[12]$ and
(Right)  $(2,3,10,15)[30]$.}
\begin{tabular}{lr}
\parbox{0.49\linewidth}{
{\scriptsize
\begin{eqnarray}                                
\alpha_{a11}&=& 1/11\left 
(11 e_1-10 e_2+e_3+e_4+...+e_{24}\right )  \nonumber\\
\alpha_{a10}&=& e_2-e_3                                 \nonumber\\
 &\cdots & \nonumber \\
\alpha_{a1} &=& e_{11}-e_{12}                               \nonumber\\
\alpha_{b11}&=& 1/11(-11 e_{25}+10 e_{24}-e_{23}-e_{22}-...-e_2)  \nonumber\\
\alpha_{b10}&=& e_{23}-e_{24}                                \nonumber\\
&\cdots & \nonumber \\
\alpha_{b1} &=& e_{14}-e_{15}                                \nonumber\\
\alpha_{c2} &=& 1/2\left ( e_{13}+e_{26}\right )-1/4 \left ( e_{12}+e_{11}+...+e_1\right )+\nonumber\\
& & +1/4 \left (e_{14}+e_{15}+...+e_{25}\right ) 
                          \nonumber\\
\alpha_{c1} &=& 1/2\left (e_{13}+e_{26}\right )+1/4\left (e_{12}+e_{11}+...+e_1\right )-\nonumber\\
& &-1/4\left (e_{14}+e_{15}+...+e_{25}\right )                           \nonumber\\
\alpha_{d1} &=& e_{13}-e_{26}                                \nonumber\\
\alpha_{0}  &=& e_{12}-e_{13}-e_{14}                            \nonumber
\end{eqnarray}
}}
&
\parbox{0.49\linewidth}{
{\scriptsize
\begin{eqnarray}  
\alpha_{a14}&=& e_{22}-e_{23}                             \nonumber\\
&\cdots & \nonumber \\
\alpha_{a1} &=& e_9-e_{10}                              \nonumber\\
\alpha_{b9} &=& -1/2\left (e_4-e_3\right )+1/4\left (e_2+e_1+...+e_8\right )
+\nonumber \\
& &+1/4\left (e_9+...+e_{23}\right )   \nonumber\\
\alpha_{b8} &=& e_7-e_8                               \nonumber\\
&\cdots & \nonumber \\
\alpha_{b1} &=& e_2-e_1                               \nonumber\\
\alpha_{c2} &=& 1/6\left (7 e_8+e_7+...+e_1+e_2-e_9-e_{10}-e_{23}\right )   \nonumber\\
\alpha_{c1} &=& e_4-e_3                               \nonumber\\
\alpha_{d1} &=& e_4+e_3                               \nonumber\\
\alpha_{0}  &=& e_3-e_2-e_9                            \nonumber
\end{eqnarray}
}
}
\end{tabular}
}

\TABLE{\caption{ 
Roots for the   $(1,6,14,21)[30]$ Berger case.}
\parbox{\linewidth}{
{\small
\begin{eqnarray}                                
\alpha_{a41} &=& e_{49}-e_{50}                            \nonumber\\
&\ldots&  \nonumber\\
\alpha_{a1} &=& e_9-e_{10}                              \nonumber\\
\alpha_{b6} &=& 1/6(5e_5-e_4-e_3...-e_2-e_1+e_9+e_{10}+...+e_{50})      \nonumber\\
\alpha_{b5} &=& e_1-e_5                               \nonumber\\
\alpha_{b4} &=& e_1-e_5                               \nonumber\\
\alpha_{b3} &=& 1/3(2e_5-e_1+e_7+e_8+...+e_{18})           \nonumber\\
\alpha_{b2} &=& e_1-e_5                               \nonumber\\
\alpha_{b1} &=& e_2-e_1                               \nonumber\\
\alpha_{c2} &=& -1/4(5e_4+e_3+e_2+e_1+e_2+...+e_5)      \nonumber\\
\alpha_{c1} &=& e_4-e_3                               \nonumber\\
\alpha_{d1} &=& -1/2(e_4+e_3-e_2-e_1-...-e_5)       \nonumber\\
\alpha_{0}  &=& e_3-e_2-e_9                            \nonumber
\end{eqnarray}
}
}}

\section{Summary, additional comments and conclusions}

The interest to look for new algebras  beyond Lie algebras started 
from   the $SU(2)$- conformal theories 
(see for example \cite{CIZ,FZ}).
One can think that  geometrical concepts, in particular algebraic geometry,
 could be a  natural and more promising way to  do this.
This marriage of algebra and geometry has been useful in both ways. 
Let us remind that to prove mirror symmetry of Calabi-Yau spaces, 
the greatest  progress was reached with using the techniques of Newton reflexive polyhedra  in Ref.\cite{Bat}.


In this work we have continued the  study of the structure of
graphs obtained from $CY_3$ reflexive polyhedra focusing on the 
description  of fourteen  ``simply laced'' cases, those graphs 
obtained from three dimensional spaces with K3 fibers which lead 
to symmetric matrices. 
We have studied both the affine and, derived from them, 
non-affine cases. We have presented root and weight structurea for them.  
We have studied in  particular those graphs leading to generalizations
of the exceptional simply laced cases $E_{6,7,8}$ and $E_{6,7,8}^{(1)}$. 
The graphs and matrices
of these simply laced graphs, both, those  already known 
of dimension 1,2,3 and those new of dimension 4 share a number of 
simple characteristics.
The cases of dimension 1,2,3 are well known and correspond to the classical
Cartan Lie algebras. 
The main objective of this work has been to enlarge this list with graphs obtained 
by vectors of dimension four (corresponding to CY3).
In dim=4, corresponding to K3-sliced $CY_3$ spaces, we have singled out by inspection the 
 following 14 RW-reflexive vectors from the total of 95- K3-vectors
{(1111),(1122), (1113), (1124), (2334),(1344),(1236),
(1225),(14510),
(1146),(1269),(1,3,8,12),(2,3,10,15)(1,6,14,21)}. 
Coxeter numbers can be  assigned in a consistent way both geometrically and algebraically.
Genuine Berger matrices are assigned to them with specially simple properties: 
they are symmetric and affine.
In addition, each of these graphs and matrices seems not 
be ``extendable'': in contradistinction to other cases \cite{volemi}, 
no other graphs and Berger matrices can be obtained from them 
simply adding more nodes to any of the legs. In this sense, these graphs are 
``exceptional''. As with the classical exceptional graphs, series can be 
traced among them. Apparently these fourteen vectors are the only ones from the 
the total of 95 vectors which lead to this kind of symmetric matrices.


It is very  well known, by the Serre theorem,
  that Dynkin diagrams defines  one-to-one   Cartan matrices and these ones Lie or Kac-Moody 
algebras.
In this work, we have generalized    
 some of the properties of Cartan matrices 
for Cartan-Lie and Kac-Moody algebras  into a new class of affine, and non-affine Berger matrices.
We arrive then to the obvious conclusion that 
any algebraic structure emerging from this 
can not be a CLA or KMA  algebra. 
The main difference of these matrices with respect 
previous definitions being in the values that 
diagonal elements of the matrices can take. In Calabi-Yau CY3 spaces, 
new entries with  norm equal to 3 are allowed.
The choice of this number can be related to two facts:  
First, we should take  in mind that in higher 
dimensional Calabi-Yau spaces resolution of singularities should be 
accomplished by more topologically complicated projective spaces:
while for resolution of 
quotient singularities in K3 case one should use the $CP^1$ with Euler 
number 2,
the, Euler number 3,  $CP^2$ space  could 
be used for the resolution of singularities in $CY_3$ space in a non-irreducible way. 
The second fact is related to the 
cubic matrix theory\cite{Kerner}, where a ternary operation is defined and 
in which the $S_3$ group  naturally appears. 
One conjecture, draft from the fact of the underlying UCYA construction, 
is that, as  
Lie and affine Kac-Moody algebras are based  on 
a binary composition law; 
the emerging picture from the consideration of 
these graphs could lead us to algebras  
including simultaneously different n-ary 
composition rules.
Of course, the underlying UCYA construction could manifest in other 
ways: for example in giving a  framework for a higher level linking 
of algebraic structures: Kac-Moody algebras among themselves and with
any other hypothetical algebra generalizing them.
Thus, putting together  UCYA theory and graphs from reflexive polyhedra,
we expect that 
iterative application of 
non-associative n-ary operations 
give us  not only a complete picture  of the RWV, but allow us in 
addition to establish
``dynamical'' links among
 RWV vector  and graphs  of  different dimensions and, in a further 
step,
links between  singularity blow-up and possibly new
generalized physical
symmetries.

\vspace{0.6cm}
{\bf Acknowledgments}.
One of us, {G.G.}, would like to give his thanks to E. Alvarez, 
P. Auranche, R. Coquereaux, N. Costa, C. Gomez, B. Gavela, L. Fellin, 
 A. Liparteliani, 
L.Lipatov,  A. Sabio Vera, J. Sanchez Solano, I. Antoniadis,
P.Sorba and  G. Valente  
for valuable discussion and nice support.
G.G. wish to acknowledge with special gratitude the 
support of the PNPI ( Gatchina, St 
Petersburg).
We  acknowledge the  financial  support of 
 the  Spanish CYCIT  funding agency (Ministerio de Ciencia y Tecnologia)
  and the CERN 
Theoretical Division. 
E.T. wish to acknowledge the kind hospitality of the 
Dept. of Theoretical Physics, C-XI of the U. Autonoma de Madrid.

{

}

\end{document}